\documentclass[prl,aps,superscriptaddress]{revtex4}

\usepackage{amsmath,amssymb,latexsym,revsymb,color,hyperref}
\usepackage[pdftex]{graphicx}
\pdfoutput=1
\usepackage{amsthm}

\newcommand{\ket}[1]{|#1\rangle}

\newcommand{\R}{\mathbb{R}}
\newcommand{\C}{\mathbb{C}}
\newcommand{\Id}{\mathbb{I}}
\newcommand{\G}{\mathcal{G}}
\newcommand{\Tr}{\operatorname{Tr}}
\newcommand{\p}{\mathcal{P}}
\newcommand{\tr}{{\rm tr}}

\newtheorem{theorem}{Theorem}
\newtheorem{definition}[theorem]{Definition}
\newtheorem{lemma}[theorem]{Lemma}

\newtheorem{corollary}[theorem]{Corollary}

\begin{document}
\title{The black hole information problem beyond quantum theory}
\author{Markus P.\ M\"uller}
\affiliation{Perimeter Institute for Theoretical Physics, 31 Caroline Street North, Waterloo, ON N2L 2Y5, Canada}
\author{Jonathan Oppenheim}
\affiliation{University College of London, Department of Physics \& Astronomy, London, WC1E 6BT and London Interdisciplinary Network for Quantum Science}
\author{Oscar C.\ O.\ Dahlsten}
\affiliation{Department of Physics, University of Oxford, Clarendon Laboratory, Oxford, OX1 3PU, UK}
\affiliation{Center for Quantum Technology, National University of Singapore, Singapore}

\date{August 18, 2012}
 
\begin{abstract}
The origin of black hole entropy and the black hole information problem provide important clues for trying to piece together a quantum theory of gravity.  Thus far, discussions on this topic have mostly assumed that in a consistent theory of gravity and quantum mechanics, quantum theory will be unmodified.  Here, we examine the black hole information problem in the context of generalisations of quantum theory. In particular, we examine black holes in the setting of generalised probabilistic theories, in which quantum theory and classical probability theory are special cases. We compute the  
time it takes information to escape a black hole, assuming that information is preserved. We find that under some very general assumptions, the arguments of Page (that information should escape the black hole after half the Hawking photons have
been emitted), and the black-hole mirror result of Hayden and Preskill (that information can escape quickly) need to be modified. The modification is determined entirely by what we call the Wootters-Hardy parameter associated with a theory. 
We find that although the information leaves the black hole after enough photons have been emitted, it is fairly generic that it fails to appear outside the black hole at this point -- something impossible in quantum theory due to the no-hiding theorem.  The information is neither inside the black hole, nor outside it, but is delocalised. Our central technical result is an information decoupling theorem which holds in the generalised probabilistic framework.
\end{abstract}

\maketitle
\setcounter{secnumdepth}{3} 

One of the central projects of theoretical physics is to construct a theory in which both gravity and quantum mechanics can be combined consistently.
In doing so, we have very little to guide us in the way of experiments.  However, the black hole does provide a model system which can aid us
in this task.  From the work of Bekenstein and Hawking, black holes possess a thermodynamical entropy which is usually attributed
to quantum gravitational microstates.  A theory of quantum gravity ought to pass the test of predicting that the entropy of a black hole is given
by a quarter of the black hole's area~\cite{HawkingOriginal,Bekenstein-entropy}.  
Likewise, a theory of quantum gravity must navigate its way through the black hole information 
problem~\cite{hawking-bhinfoloss,hawking-unpredictability,preskill-infoloss-note}.  Namely, either the theory
preserves information (as in quantum theory), in which case it ought to explain how information apparently is able to escape the black hole horizon.  
Or, if information is destroyed, it must explain how this can happen while still apparently preserving conservation laws~\cite{bps,unruh-wald-onbps,OR-intrinsic}.

However, thus far, the discussion on the black hole information problem has largely been within the context of theories in which quantum 
mechanics is unmodified.  This seems an undue restriction, given that the central motivation for studying black hole thermodynamics and information
is that it can lead us to other consistent theories which are experimentally compatible with gravity and quantum theory.  If studying black holes is 
going to allow us to explore what form a theory of quantum gravity might take, then we shouldn't be confining ourselves to theories in which quantum theory is unmodified.
The aim of the present article is to begin to remedy this, by expanding the discussion to include generalisations of quantum theory, 
in the hope that it will allow us to explore black hole information in a more robust setting.

We will take as our starting point the assumption that the consistent theory of nature fits in the class of {\it generalised probabilistic
theories} (GPTs). This is a very unrestrictive framework.
Quantum theory and classical theory are but special cases of GPTs. Crucially, any theory whose operational output is the probabilities
of outcomes of measurements, conditional on a choice of system preparation and subsequent transform, can be formulated in this way~\cite{Hardy}.
GPTs also have a natural notion of reversible time evolution, generalising the unitary time evolution of quantum mechanics.
GPTs are currently being studied extensively in quantum information theory, since examples of GPTs exist that exhibit interesting
information-theoretic behaviour that deviates from quantum theory, such as superstrong nonlocality~\cite{Khalfin,Tsirelson,PopescuRohrlich,Barrett07}.
We recommend~\cite{Hardy, Barrett07,Mielnik74} as background for readers unfamiliar with GPTs.

We will take as an axiom that information is preserved in such theories in that time evolution is reversible
(akin to unitarity in quantum theory), and then use this framework to study information in black holes.  In particular, what is of interest is the tension between information preservation and Hawking's calculation, which suggests that at least semi-classically, a black hole radiates information thermally, apparently destroying information.  If information
is preserved, and escapes from a black hole before quantum gravitational effects come into play, 
then one can find a set of space-like hypersurfaces, such that information appears to be cloned~\cite{susskind-thorlacius-note-preskill}. Cloning is not possible in quantum theory or any other generalised probabilistic theory apart from classical probability theory~\cite{BarnumBLW06}.  On the other hand, if Hawking's semi-classical calculation holds until the black hole is of the Planck mass, 
and quantum gravitational effects come into play, then one effectively has a long lived black hole remnant with a lot of information stored inside, and this presents a host of associated difficulties~\cite{preskill-infoloss-note,tHooft-bhcompl,acn87,CarlitzWilley}.

The speed at which information leaves the black hole is thus an important question. In quantum theory, three cases of particular interest here have been considered:  (i) for the case of a black hole which is initially
in a pure quantum state, Page argued that if information is preserved, then, under certain assumptions, it would have to start escaping when half the photons had been emitted~\cite{page-unitary-evap}, and thus the theory needs to find a way around the cloning argument. Black hole complementarity~\cite{tHooft-bhcompl,tHooft-bhcompl-string,susskind-bhcompl} is one such mechanism to avoid the cloning argument, (ii) classical information on the other hand, can be locked inside a black hole until the very end of the evaporation process~\cite{bhlock} without suffering from the problems usually associated with remnants, (iii) the same mechanism which produces the locking of the classical information,
will also cause a black hole to emit its information almost instantly, as if it is a mirror, in the case where the state of the black hole is initially entangled with the outside~\cite{HaydenPreskill}.  This later result pushes any mechanism which avoids the cloning argument to its very limits, and has possible
implications for the amount of time it takes for black holes to scramble information~\cite{braunsteinciphers,sekino2008fast}.

In this paper, we study scenarios (i) and (iii) in the general setting beyond quantum theory. We compute the corresponding information retrieval properties for all GPTs that satisfy some natural assumptions, including quantum and classical probability theory as special cases. It turns out that post-quantum theories behave
quantitatively different from quantum theory: in Page's scenario (i), the black hole may emit much more than half of the photons until information escapes. Hayden and Preskill's mirror result for case (iii) remains valid qualitatively, but with an interesting difference. To analyze this result in the generalized context, we prove a version
of the \emph{decoupling theorem}~\cite{braunsteinciphers,how-merge2,fqsw} for GPTs.  A decoupling theorem essentially tells us how easy it is to remove the correlations from
a state. In quantum theory, the decoupling theorem would tell us how quickly information generically leaves a black hole, and also, how quickly this information appears
outside the black hole. This is because in quantum theory, information cannot be encoded in the correlations between the black hole and the radiation,
but must reside almost entirely in one or the other. Thus, if information cannot be found inside the black hole any more, 
it must be localised outside of it.  In the context of quantum information theory, this is a consequence of decoupling~\cite{how-merge2}, and
in the context of black holes, it is the {\it no-hiding theorem}~\cite{braunstein-pati-nohiding}). Its approximate version is related to
Uhlmann's theorem~\cite{uhlmann1976transition} (see \cite{how-merge2}).

We find that for GPTs, the mirror results remains
valid, but the no-hiding theorem does not hold in general. This leads to the intriguing possibility that information can escape the black hole quickly, but not be found outside of it,
instead becoming delocalised.  If the information is delocalised for a long enough time, then this could 
potentially serve as an alternative to black-hole complementarity, as it avoids the problem of there being a hypersurface in which there is a 
copy of the information both inside and outside the black hole.

We start by first describing the general class of theories we consider, and then the physical situation of black hole evaporation 
as recast in \cite{bhlock} for the purpose of an
information-theoretic analysis. Our central technical result is Theorem \ref{TheDecoup}, proven in the Appendix.  After introducing it, we apply it to scenarios (i) and (iii) above and contrast our results for post-quantum theories to the known quantum results.  Our main conclusion is that for generic
potential generalisations of quantum theory one can have preservation of information, but the rate at which information 
leaves the black hole is modified.  In particular, information can escape very late in the evaporation process, and can even be delayed until the point
when the black hole is no longer semi-classical, thus respecting the semi-classical result of Hawking, yet without resulting in the problems associated
with information crossing a causal horizon.  Likewise, the fact that information can become delocalised in such theories could potentially be used
as an alternative to black-hole complementarity.   

\section{General probabilistic theories}
 
In General Probabilistic Theories (GPTs), one assigns states $\omega$ to any physical system (in quantum theory, this would be the density matrix $\rho$). If $A$ is a physical system, the set of all states, the \emph{state space}, will be denoted $\Omega_A$; it can always be chosen as some subset of $\R^n$ with suitable $n$. By assumption, it is possible to prepare either some state $\omega$ with probability $p$, or some state $\varphi$ with probability $1-p$, yielding $p\omega+(1-p)\varphi$.
Thus, every state space $\Omega_A$ is convex, and for further physical reasons compact.
This is also true for quantum $n$-level systems, where the state space is the convex set of $n\times n$ density matrices. Similarly as in quantum theory, we call a state \emph{mixed} if it can be written in the form $p\omega+(1-p)\varphi$ for some $0<p<1$ and $\omega\neq \varphi$, and otherwise \emph{pure}.
Thus far, (GPTs) have only been studied in the context of describing a physical system which exists in space-time -- they have not been applied to 
the description of space-time itself.  While it is likely that a proper account of black hole information will need to concern itself with describing
space-time, such considerations are clearly beyond our current understanding.

Imagine some measurement with $k$ outcomes. Applying it to some state $\omega$, the probability to obtain specifically the first outcome can be denoted $e_1(\omega)$,
which is a real number in the interval $[0,1]$. If we feed a statistical mixture into the measurement device, we get the probability $e_1(p\omega+(1-p)\varphi)=p e_1(\omega)
+(1-p)e_1(\varphi)$. Thus, $e_1$ is a linear map which is non-negative on all states -- we call these maps \emph{effects}. The further measurement outcomes
are similarly described by effects $e_2,\ldots, e_k$ such that the total probability is $\sum_i e_i(\omega)=1$. In quantum theory, if $\omega$ is a density matrix,
every effect $e$ has the form $e(\omega)=\tr(P\omega)$, with some matrix $0\leq P \leq \mathbf{1}$ (e.g.\ a projector).

Transformations must map states to states -- since they must respect statistical mixtures, they must be linear. In the following, we are only interested in \emph{reversible
transformations} $T$, that is, ones that have an inverse transformation $T^{-1}$ and thus do not destroy information. We do not consider
transformations which destroy information, since the
entire crux of the black hole information problem is the question of whether information preserving transformations are consistent with what we know about black holes.  To every physical system $A$, there is a compact (possibly finite)
group of reversible transformations $\G_A$. In quantum theory, these are the unitaries, $\rho\mapsto U\rho U^\dagger$. 
They are symmetries of the state space.

\begin{figure}[!hbt]
 \begin{center}
  \includegraphics[angle=0, width=5cm]{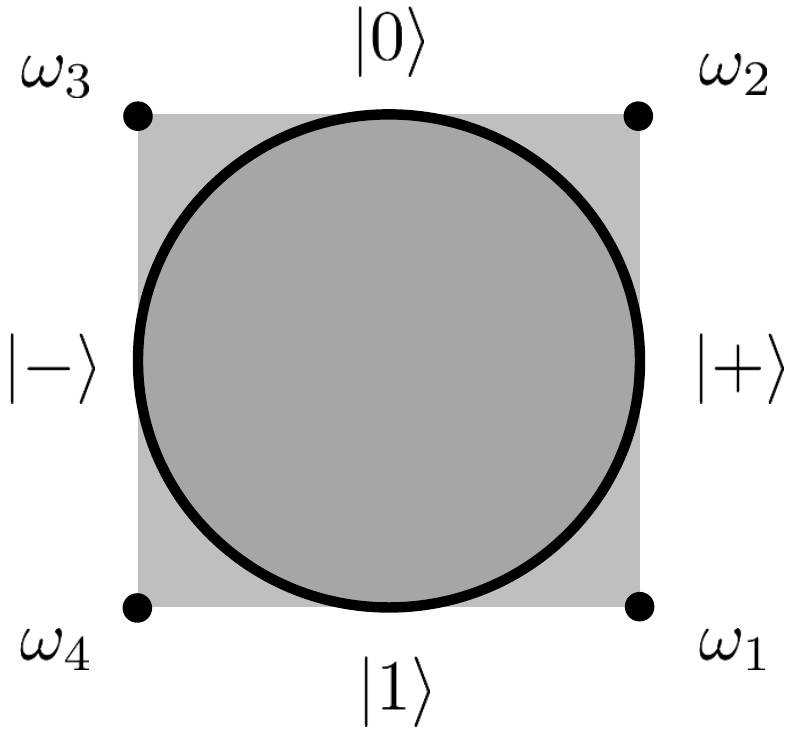}
 \caption{A simple example of GPT state spaces other than quantum theory. The inner circle is the equatorial plane of the quantum theory Bloch sphere, with all states on the circle pure $(\ket{\pm}:=\frac{1}{\sqrt{2}}(\ket{0}\pm\ket{1}))$. One may alternatively consider the outer square as the state space, in which case there are only four pure states, $\omega_1,\ldots,\omega_4$. As in any theory in the GPT framework, any convex combination of states is allowed, implying that any point in the square must be an allowed state. This outer square state space, which contains the quantum states, is a non-quantum example of a GPT state space, and is called a `gbit'~\cite{Barrett07}. The only possible reversible transformations
 would be rotations by multiples of $90^\circ$ and reflections across the center.}
 \label{fig_square}
 \end{center}
 \end{figure}

Figure~\ref{fig_square} gives an example of a GPT state space other than quantum theory~\cite{Barrett07}. 

In the following calculations, it turns out that two quantities will be of paramount importance that have first been introduced by Wootters and Hardy~\cite{Wootters,Hardy}.
Given some system $A$, we denote by $K_A$ the dimension of the set of unnormalized states; that is, $K_A=\dim(\Omega_A)+1$, because $\Omega_A$ is the set
of normalized states. Furthermore, we denote by $N_A$ the maximal number of states that are perfectly distinguishable in a single measurement. 

If $A$ is a quantum
$n$-level system, then we can perfectly distinguish at most $n$ (orthogonal) states, hence $N_A=n$. In general, states $\omega_1,\ldots,\omega_n$ are perfectly
distinguishable if there is an $n$-outcome measurement with effects $e_1,\ldots, e_n$ such that $e_i(\omega_j)=\delta_{ij}$.
In quantum theory, $\Omega_A$ is the set of $n\times n$-density matrices; hence $K_A=\dim(\Omega_A)+1=n^2$; this is the number of independent real parameters
in an unnormalized density matrix. In other words, in quantum theory, we have $K=N^2$. In contrast, a classical $n$-level system is described by a probability distribution
with $n$ parameters $(p_1,\ldots,p_n)$. Thus, in classical probability theory, we have $K=N$. GPTs can have arbitrary relations between $K$ and $N$; it can only be proven
in general that $K\geq N$.

GPTs also have a notion of composite systems. The quantum notions of subsystems and tensor products generalise to all GPTs under the standard
assumption that signalling is not possible, i.e.\ that the reduced state on a subsystem is invariant under local operations on other subsystems~\cite{Barrett07}.
This assumptions is implicit in the GPT framework, and also in this paper: one party cannot simply send information to another party by choosing local measurements.

\section{Our working assumptions}
\begin{figure}[!hbt]
 \begin{center}
\includegraphics[angle=0, width=12cm]{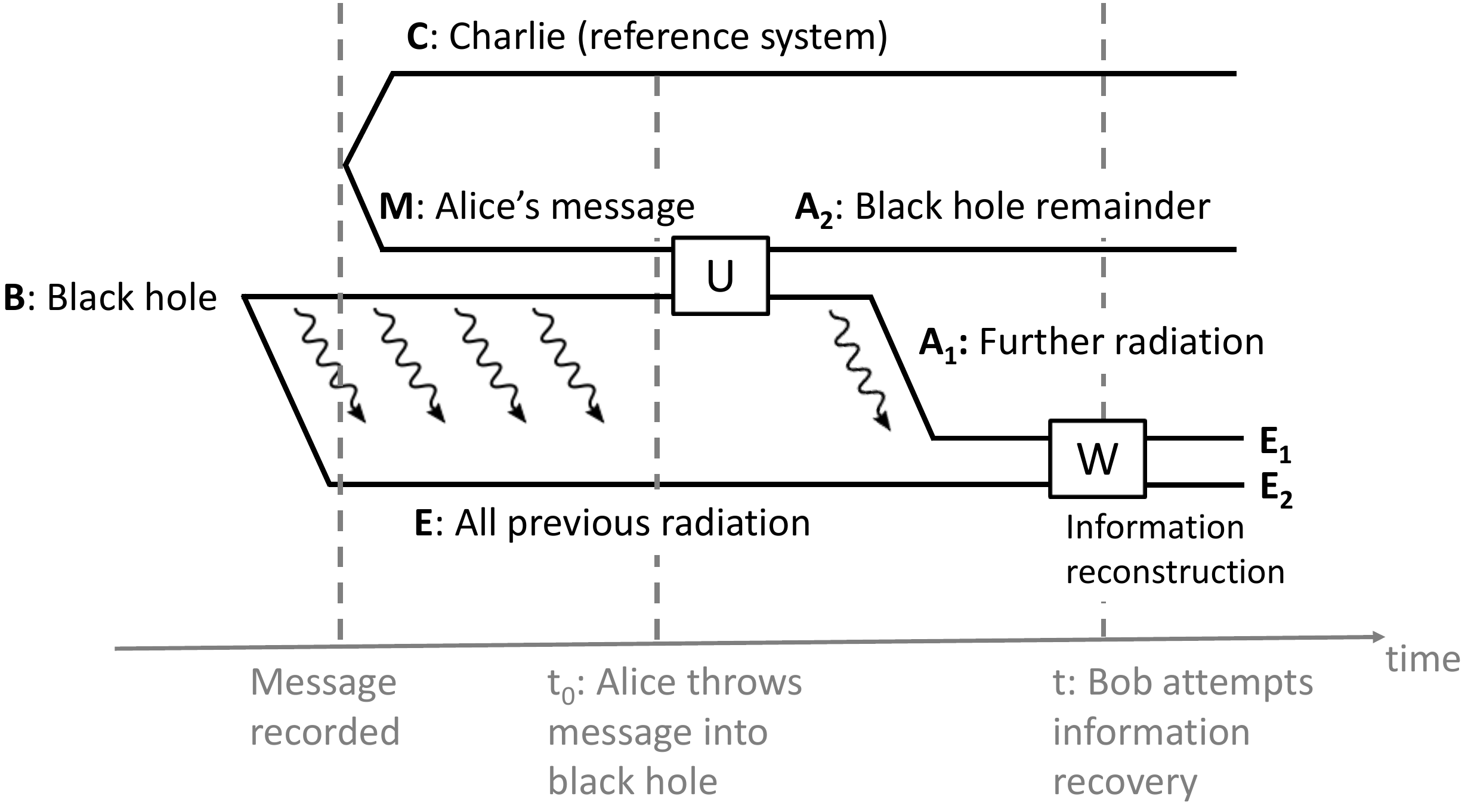} 
\caption{Alice throws her message into the black hole. There are three parties: Alice's message (M), the earlier black hole (B), Bob (E) and the reference system (C). (The reference system is particularly natural to include in the case of quantum theory where one may always take C to be a purifying system such that $\psi^{AEC}$ is pure, but we shall not be assuming such purification is always possible.) Then a reversible interaction $U$ is applied to Alice's system, representing the black hole dynamics acting on her diary after she has thrown it in. At some given subsequent time some of the black hole ($A_1$) has leaked out, e.g. via Hawking radiation, and is now in the possession of Bob, and relabelled as $E_2$. Bob also holds any radiation predating Alice's message having entered the black hole ($E$), and he can perform a joint operation $W$ on
the system $EA_1=E_1 E_2$. We also have $BM=A_1 A_2 = A$.}
 \label{fig_blackhole}
 \end{center}
 \end{figure}

We are now interested in the class of theories for which the calculations of Page, Hayden and Preskill can be meaningfully generalised. In particular, we will be interested
in the situation in Figure~\ref{fig_blackhole}, which will be explained in more detail below. This setting involves a composition of four state spaces $A=A_1 A_2$, $E$, and $C$.
In general, we can imagine that each system is described by an arbitrary GPT, with an arbitrary choice of a compact convex state space. However, it is clear that at least
some requirements on the state spaces must be satisfied such that the setting makes physically sense -- if the group of transformations $\G_A$ contained only the identity map, for example,
then no interesting dynamics could happen whatsoever.

We now describe the technical assumptions that we impose on our state spaces, together with their physical meaning.
Our first assumption is called \textbf{transitivity}: for every pair of pure states $\varphi,\omega$ in a common state space, there is a reversible transformation $T$ such that
$T\varphi=\omega$. In our context, this is a very natural assumption: in order to study the black-hole information paradox, we only consider reversible time evolution. Moreover,
we imagine that pure states are prepared by starting from a single reference state (like the vacuum state, for example) and applying some reversible time evolution.

Our second assumption is of a technical nature: we assume that the group of reversible transformation $\G_A$ acts \textbf{irreducibly} on the state space; in fact, we assume
that it is an irrep in the usual sense of group representation theory~\cite{Simon}. This assumption is not crucial -- our calculations can be done for more complicated
situations, but it keeps the calculations and results simple to start with. It is true for the group of unitary conjugations in quantum theory, and also for the state space of
classical statistics, where $\G_A$ consists of the permutations of entries of the probability vector.

Our next assumption comes from the physical requirement that \textbf{state spaces contain ``classical'' subsystems.}
That is, on every state space $A$, there should be a set of perfectly distinguishable, pure states $\omega_1,\ldots,\omega_{N_A}$ that have all the properties
of ``classical'' configurations: they can be permuted by reversible time evolution, and their uniform mixture is the state of ``maximal ignorance'', the maximally
mixed state $\mu^A$. When we have a compound system $AB$, then its classical subsystem can be obtained from combining the classical subsystems of $A$ and $B$.

In quantum theory, a classical subsystem would be the states in some orthonormal basis. The existence of classical subsystems is empirically motivated: in some limit, or,
say, after decoherence, systems behave very classically. In our setting, it makes sense to assume that the physically relevant GPT contains quantum theory as a subspace,
which in turn contains classical subsystems as states in some orthonormal basis.

Our final requirement is on how different state spaces $A$ and $B$ are combined into a joint state space $AB$. In GPTs, joint state spaces must satisfy
some minimal requirements: if $\omega^A$ and $\omega^B$ are states on $A$ and $B$, there must always be a ``product state'' $\omega^A\omega^B$ on $AB$,
and similarly for effects and transformations. This already implies for the dimensions that $K_{AB}\geq K_A K_B$. We now assume that $K_{AB}=K_A K_B$ -- that is,
that the number of degrees of freedom of the joint state space is in this sense ``minimal''. In our setting, this is a very natural and almost
mandatory assumption: if $K_{AB}>K_A K_B$, then there are holistic degree of freedoms which are neither localised in the black hole nor outside of it.
Information inside the black hole could then be transferred to these extra holistic degrees of freedom which could then only be accessed by joint measurements on the black hole and systems outside it. This is perhaps an interesting potential route to allowing information to be preserved as the system originally carrying it enters the black hole. However the question of how quickly information leaves the black hole is not well-defined or perhaps even relevant then and we shall therefore not discuss this case further here. 

The assumption that  $K_{AB}=K_A K_B$ has an information-theoretic interpretation that is sometimes called \textbf{local tomography}: every state $\omega^{AB}$ on $AB$ is uniquely
determined by the statistics and correlations of local measurements on $A$ and $B$. In other words, to determine a global state $\omega^{AB}$, it is sufficient to perform local measurements and subsequently analyze the correlations of local outcomes (obtained from many measurements on independent copies).
It can also be rephrased as the fact that the product states $\omega^A\omega^B$ span the composite state space. By our definition of $K_A$, the unnormalized
states are vectors in the real linear space $L_A:=\R^{K_A}$. So local tomography means that $L_{AB}=L_A\otimes L_B$ -- that is, unnormalized global states
are carried by a vector space that is the tensor product of the local vector spaces. Again, this requirement holds for classical and quantum state space, and it is
also true for most alternative GPTs that have been studied in the literature~\cite{Barrett07}.

We call the above requirements the ``standard assumptions''; see Definition~\ref{DefStandardAssumptions} in the appendix for the full mathematical details.
They imply also that $N_{AB}=N_A N_B$.

We additionally require that although the fundamental theory of nature may not be quantum theory, physics outside the black hole is very
well described by quantum theory,
and that semi-classical gravity also remains valid. So for example, systems inside the black hole and at Planck energy, can behave very differently to systems in quantum theory, but
the photons which escape the black hole should obey quantum theory. This is related to our ``classical subsystems''-assumption above: states that describe systems
far outside the black hole should lie in a quantum ``subspace'' of the more fundamental GPT, similarly as classical systems can be thought of as occupying a subspace of quantum theory (given
by diagonal density matrices).

\section{The physical setup}
We now fix some notation and describe the physical situation.
At some time $t_0$, Alice holds the state of system $M$.  We will assume that this state can be described by quantum theory, since it is outside the black hole.
The state is entangled with an external referee $C$
which we call Charlie. That is, there is a global entangled state $\psi^{MC}$ which is held by Alice and Charlie, who are both outside of the black hole;
to simplify the calculation we assume that it is pure.
Essentially, we can interpret the correlation between $M$ and $C$ as meaning that system $M$ has information about $C$.  This will
allow us to quantify what it means for information to be inside the black hole, or to escape from the black hole.

Then, Alice throws her system which is in some (mixed) marginal state $\psi^M$ into a black hole $B$. The black hole was formed
in the far past in a pure state; since then, it has already potentially emitted Hawking radiation. The subsystem carrying all previously emitted Hawking radiation
is denoted $E$.
Alice does not take part in the rest of this thought experiment. We denote the joint system $BM$ by $A$, and the total initial state at time $t_0$
by $\psi=\psi^{CAE}=\psi^{CMBE}$.   In what follows, unless otherwise indicated, all states and transformations are in the context of GPTs.

We now consider some later time $t>t_0$, during which the black hole has been evaporating. Note that we use the notion of ``time'' merely as an illustration,
and not as an ingredient in actual calculations: all that is important for our setup is the ordering in which different subsystems are held by different parties.
For concreteness, we may imagine that $t$ is the time measured by the outside observer Charlie. The quantity that is relevant for our final result turns out to be
the number of emitted radiation quanta, $\log N_{A_1}$.

We assume that black hole evaporation
is accomplished by some total reversible time evolution $U$. The input of $U$ is the system inside the black hole, i.e.\
the original black hole $B$ and the system $M$ that Alice subsequently threw into it. That is, $U$
is some reversible transformation acting on $A=BM$.
Additionally, some of what was in the black hole is emitted as Hawking radiation.
The system composed of all quanta that have been emitted between times $t_0$ and $t$
is denoted by $A_1$; we denote by $A_2$
the part which remains inside the black hole until time $t$. As state spaces, we have $A=A_1A_2=BM$.
The action of $U$ can be understood as randomly selecting a subsystem $A_1$ to be emitted, which is very
similar to the setup in~\cite{braunsteinwithoutspacetime}. Denote the global state
at time $t$ by $\sigma(U)$ (since it depends on $U$). That is, $\sigma(U)^{CAE}=U_A\otimes\Id_{CE} (\psi^{CAE})$
since the system $CE$ is far from the black hole, and so remains unchanged (our calculations remain valid if these systems evolve
reversibly locally in a way which does not create correlations with the other parties).

Recall the state space $E$ associated to the previously emitted Hawking radiation; in our thought experiment, we imagine
that all the emitted quanta have been collected by an agent that we call Bob who resides far away from the black hole.
At time $t$, this will also include all the radiation which has been emitted from the black hole up to this point, and
thus we imagine that Bob has access to the joint system $EA_1$. At this point, Bob may attempt to decode some of Alice's
information by applying an arbitrary local transformation $W$; this way, he may hope to obtain (some of) the correlations with Charlie that
were initially held by Alice.

\section{The decoupling theorem for GPTs}
We are now interested in knowing when the information that Alice has put into the black hole leaves it.  That is, we would like to know when the
system $A_2$ which remains inside the black hole is decoupled from the reference system $C$, such that $\sigma(U)^{CA_2}\approx \psi^C\otimes \mu^{A_2}$, 
where $\mu^{A_2}$ denotes the maximally mixed state on $A_2$. In particular, we are interested in how much needs to be removed (i.e. how big $A_1$ needs to be) until there are no correlations between what remains inside the black hole ($A_2$) and the reference system $C$ that $A$ was originally correlated to. 
 Intuitively, a statement like this is important in the context of the black hole information problem, because it can tell us how many photons typically escape the black hole, 
before information leaks out. This is because information is always information {\it about something}, and in this case,
the correlation between $A$ and $C$ can be thought of as the information that $A$ has about $C$.  In other words, we think of $C$ as being
the source of the original information, and
$M$ is information about $C$. So, if initially $A$ and $C$ are fully correlated, and then
after some time $t$, the part $A_2$ which is inside the black hole is no longer correlated with $C$, then we know that all the information has left the black hole.
In the quantum case, this also implies that the information is now outside the black hole, but we will see that this is not the case for more general theories.

If the total system is quantum, the standard decoupling theorem says the following.
For almost all evolution laws~\footnote{By this we mean that if we select a unitary at random, from the uniform measure,
then with high probability the statement holds.} this property holds: $\sigma(U)^{CA_2}\approx \psi^C\otimes \mu^{A_2}$, 
where $\mu^{A_2}=\mathbf{1}/d_{A_2}$ denotes the maximally mixed state on $A_2$, provided that the dimension $d_{A_1}$ of the part $A_1$ of $A$ which is removed satisfies
\begin{align}
2\log{d_{A_1}}\gg\log{d_A}+\log{d_C}-\log\tr\left((\psi^{CA})^2\right).
\label{eq:q-decoupling}
\end{align} 
 That is, the remaining black hole $A_2$ is almost uncorrelated with (``decoupled'' from) $C$, provided that enough radiation has emerged
from the black hole.   In many cases, Equation \eqref{eq:q-decoupling} implies that $2\log{d_{A_1}}\gg n\cdot I(C:A)$, where 
$I(C:A):=S(C)+S(A)-S(CA)$ is the mutual information, with $S(X):=-\tr(\rho_X\log{\rho_X})$ the von Neumann entropy.

Furthermore, in the quantum case, the fact that information has left the black hole (i.e.\ the black hole has decoupled from $C$), implies
that the information now is located outside the black hole, and can be reconstructed by Bob~\cite{how-merge2}.  This is because in quantum theory, up to a unitary
on the purifying system, there is only one pure state compatible with any density matrix on $AC$.  So if the remaining state
of the reference system and the black hole is decoupled, i.e. $ \psi^C\otimes \mu^{A_2}$, then there exists an isometry $W$ acting on the purifying system $E$
and taking it to systems $E_1\simeq M$ and $E_2$ such that the state on the entire system is
$(W^E\otimes{\rm Id}^{CA_2})\left(\strut\sigma(U)\right)\approx \tilde\psi^{CE_1}\otimes\phi_+^{E_2 A_2}$,
where $\tilde\psi^{CE_1}\approx \psi^{CM}$, and $\phi_+$ denotes a maximally entangled state. That is, the correlations with $C$ that were initially
located in system $M$ are now on Bob's system $E_1$.
It is this general principle (related to Uhlmann's theorem) which
has been dubbed ``no-hiding'' in the context of black holes~\cite{braunstein-pati-nohiding}.

We now wish to explore how this result becomes modified if our `quantum' theory of gravity involves a GPT other than quantum theory.

Below we state a {\bf general probabilistic decoupling theorem}, proven in the appendix. We need to introduce two notions before stating the result.  Firstly, the {\em $2$-norm}
$\|\cdot\|_2$ is the usual Euclidean norm with the subtlety that the state space is represented such that all reversible transformations are orthogonal
while all pure states are on the unit sphere, surrounding the maximally mixed state. The generalised {\em purity} $\p(\psi)$ is  defined as $\|\psi-\mu\|_2^2$,
where $\mu$ is the maximally mixed state. This is
$1$ if and only if $\psi$ is pure (not a mixture of other states), and $0$ if and only if $\psi$ is the maximally mixed state~\cite{TypEnt}.
See Table~\ref{TableComparison} in the appendix for what these notions are in the special cases of quantum theory and classical probability.

\begin{theorem}[Decoupling in GPTs]
\label{TheDecoup}
Consider the situation depicted in Figure~\ref{fig_blackhole}, with the notation $\sigma(U):=U_A\otimes\Id_{CE}(\psi)$, where
$U\in\G_A$ is a random reversible transformation. If the standard assumptions hold, we have
\begin{eqnarray*}
\int_{U\in\G_A} \left\| \sigma(U)^{CA_2}-\psi^C\otimes\mu^{A_2}\right\|_2^2\, dU=
\p(\psi^{CA})\cdot\frac{(K_{A_2}-1)(N_C N_A-1)}{(N_C N_{A_2}-1)(K_A-1)}
-\p(\psi^C)\cdot\frac{(N_C-1)(K_{A_2}-1)}{(N_C N_{A_2}-1)(K_A-1)}.
\end{eqnarray*}
Under an extra assumption on the subsystem $CA_2$ (cf.\ Theorem~\ref{TheDec1} in the supplementary material), we get
\begin{equation}
   \int_{U\in\G_A} \|\sigma(U)^{CA_2}-\psi^C\otimes\mu^{A_2}\|_1^2\, dU \leq \p(\psi^{CA})\cdot\frac{N_C N_A}{K_{A_1}}
   \label{eqDecoup1}
\end{equation}
if all involved $N$ and $K$ are large, where $\|\omega-\varphi\|_1$ denotes the maximal difference in probability of any outcome of any possible
measurement that can be applied to the states $\varphi,\omega$.
\end{theorem}

The theorem is proven in the technical supplement, which also contains the detailed definitions of the GPT framework and of our assumptions.

\section{Implications for Hayden-Preskill scenario}
Similarly as Hayden and Preskill~\cite{HaydenPreskill}, we consider the situation that the black hole has been radiating for a very long time,
such that it has become maximally entangled with its Hawking radiation. The state $\psi^{BE}$ is maximally entangled at time $t_0$,
by which we mean that the reduced state $\psi^B$ is close to the maximally mixed state, $\psi^B=\mu^B$.
The state $\psi^{CA}$ which appears in the theorem above is then
\[
   \psi^{CA}=\psi^{CMB} = \psi^{CM}\otimes \mu^B
\]
with $\psi^{CM}$ pure. According to Theorem 28 in~\cite{TypEnt}, its purity is
\[
   \p(\psi^{CA})=\frac{N_C N_M-1}{N_C N_M N_B-1}\approx \frac 1 {N_B}=\frac{N_M}{N_A},
\]
where the approximation is true if all $N$'s are large enough such that subtracting unity can be neglected. According to Theorem~\ref{TheDecoup},
this gives a criterion for when the state $\sigma(U)^{CA_2}$ is operationally (i.e.\ in $\|\cdot\|_1$-norm) almost indistinguishable from the uncorrelated
state $\psi^C\otimes\mu^{A_2}$:
\[
   \sigma(U)^{CA_2}\approx \psi^C\otimes \mu^{A_2} \quad\mbox{ if }\quad \p(\psi^{CA})\cdot\frac{N_C N_A}{K_{A_1}} \ll 1
   \quad\Leftrightarrow\quad K_{A_1}\gg N_{MC}.
\]
In other words, the black hole must have radiated away ``enough'' Hawking radiation in $A_1$, with a lower bound given by the size of
the system shared between Alice ($M$) and Charlie ($C$). To simplify the discussion, we may assume that Alice and Charlie carry the same
types of systems, i.e.\ $N_M=N_C$, such that $N_{MC}=N_M^2$.

Let us first consider the case of \textbf{quantum theory}. In this case, we have $K_{A_1}=N_{A_1}^2$, and so the decoupling condition becomes
$N_{A_1}\gg N_M$. If Alice's state consists of $k$ qubits, we have $N_M=2^k$, and the condition is ensured if $k+c$ qubits have been radiated away
in $A_1$ (with some small constant $c$), because then $N_{A_1}=2^{k+c}\gg N_M$. We have thus recovered the result by Hayden and Preskill: if the
black hole has radiated away enough of its degrees of freedom such that it is maximally entangled with the Hawking radiation, then it acts as a ``mirror'',
effectively ``bouncing back'' in just a few additional qubits any quantum information that is thrown in.

Theorem~\ref{TheDecoup} and the discussion so far have not touched on the question how the outside agent Bob can actually obtain the information
that the black hole has radiated away. As discussed above, in quantum theory, Bob may recover the information by applying a suitable unitary $W$ on his system $EA_1$.
This is guaranteed by the ``no-hiding theorem'' or Uhlmann's theorem.

Now we consider the situation \textbf{beyond quantum theory}. As explained above, any GPT appearing in the context of known physics must contain
quantum theory's state space as a subspace -- we expect that it behaves like quantum theory in non-extreme situations (like outside the black hole).
Thus, there must be enough degrees of freedom $K$ such that quantum theory with its $K=N^2$ degrees of freedom is contained; i.e.\ $K\gg N^2$.

In order to get a more concrete picture of the decoupling situation, we will assume that there is a functional relation between $N$ and $K$ of the
form $K=N^r$, where the Wootters-Hardy parameter $r\geq 3$ is an integer (we have $r=2$ in quantum theory, and $r=1$ in classical statistics). In fact, we show in Lemma~\ref{LemKNr}
in the appendix that this relationship follows from a few simple additional assumptions on top of our standard assumptions.
Thus, the decoupling condition becomes $N_{A_1}^r \gg N_M^2$, or for the number of generalized radiated bits $\log N_{A_1}$,
\[
   \sigma(U)^{CA_2}\approx \psi^C\otimes \mu^{A_2} \quad\mbox{ if }\quad  \log N_{A_1} \gg \frac 2 r \log N_M
   \qquad \mbox{ (for }N_M=N_C\mbox{ and }\psi^B\mbox{ maximally mixed)}.
\]
This shows that for theories beyond quantum theory, information can leave a black hole even faster than in the quantum case.
Surprisingly, if $r>2$, then this inequality may be satisfied even if the number of radiated bits $\log N_{A_1}$ is \emph{less}
than the number of generalized bits $\log N_M$ that Alice has thrown into the black hole. That is, the black hole gets decoupled
from Charlie even \emph{before} it had any chance to output all information that Alice has put in.

How is this possible? The only conceivable explanation seems that the no-hiding theorem of quantum theory loses its validity: even
if the black hole gets decoupled from Charlie, the outside agent Bob is still not able to obtain the correlations with Charlie by applying
any reversible transformation $W$. Otherwise, Alice's $k$ generalized bits would somehow have ended up at Bob's place by transmitting
$2k/r$, i.e.\ much less than $k$, bits.

We can interpret this finding by saying that for theories with $r>2$, information leaves the black hole more quickly than in the quantum case,
but may not become accessible outside the black hole. We can also obtain a bound on the time when the information starts to appear outside.
Suppose that $\sigma(U)^{CEA_1} \approx \psi^{CE}\otimes \mu^{A_1}$. This means that adding the $A_1$-system to the full outside world $CE$
increases the entropy of the outside system; it only adds noise. This means that the emitted system $A_1$ will not carry information to the outside
world. Again, we can use eq.~(\ref{eqDecoup1}) to obtain a bound on when this happens: checking that this formula remains valid for
the $CE$-versus-$A_1$ cut, this situation happens if $N_{CE}N_A\ll K_{A_2}$. Assuming that the black hole has previously radiated away half of its
degrees of freedom, i.e.\ $N_E=N_B$, we obtain
\[
   \sigma(U)^{CEA_1}\approx \psi^{CE}\otimes\mu^{A_1} \quad\mbox{ if }\quad \log N_{A_1}\ll \frac{r-2} r \log N_A
   \qquad\mbox{(for }N_M=N_C\mbox{ and }N_E=N_B).
\]
That is, at least $(r-2)/r\log(N_A)$ generalized bits have to be emitted before information starts to appear outside. In quantum theory $(r=2)$, this is
a trivial bound: information starts to appear directly, but not so for theories with $r\geq 3$.
Thus in this scenario, and for sufficiently large $r$, although information escapes from the black hole even more quickly,
it takes even longer for the information to appear outside the black hole.

In order to analyze the failure of the no-hiding theorem beyond quantum theory,
we now consider the \textbf{simplified situation where the system $E$ does not exist}, or, in other words,
is one-dimensional: $N_E=K_E=1$. Physically, this means that Alice throws her system into a very young black hole $B$ that has not radiated
so far (we may also imagine that she forms the black hole $A=MB$ at time $t_0$ from her state $M$ and some other massive stuff $B$).
Then the situation becomes fully symmetric with respect to interchanging $A_1$ and $A_2$. Thus, in analogy to above, we can compute bounds
on the number of radiated bits which guarantee decoupling of subsystems. We consider two scenarios:
\begin{itemize}
\item \textbf{Black hole is decoupled from Charlie.} This indicates that the information has left the black hole.  This is the case if $\sigma(U)^{CA_2}\approx \psi^C\otimes\mu^{A_2}$,
which holds due to eq.~(\ref{eqDecoup1}) if $N_C N_A\ll K_{A_1}$, that is, if
\begin{align}
   \log N_{A_1} \gg \frac 1 r \log N_A + \frac 1 r \log N_C.
\end{align}
(Here we are \emph{not} assuming that $N_M$ and $N_C$ are necessarily identical, in contrast to the beginning of this section. We are also \emph{not}
assuming that $\psi^B$ is mixed, taking into account that the black hole may not have radiated previously, and may thus still be pure.)
\item \textbf{Radiation is decoupled from Charlie.} This indicates that no information has arrived outside the black hole, since the emitted radiation is completely uncorrelated with the original
source of information $C$.  This is the case if $\sigma(U)^{CA_1}\approx \psi^C\otimes \mu^{A_1}$. Swapping $A_1$ and $A_2$ in eq.~(\ref{eqDecoup1}),
we see that this holds if $N_C N_A\ll K_{A_2}$, i.e.
\begin{align}
   \log N_{A_1} \ll \frac{r-1} r \log N_A - \frac 1 r \log N_C.
\label{eq:decoupledfromrad}
\end{align}
In the following discussion we assume that $N_A\gg N_C$, i.e.\ that the black hole is much larger than Alice's system.
\end{itemize}
 \begin{figure}[!hbt]
 \begin{center}
\includegraphics[angle=0, width=11cm]{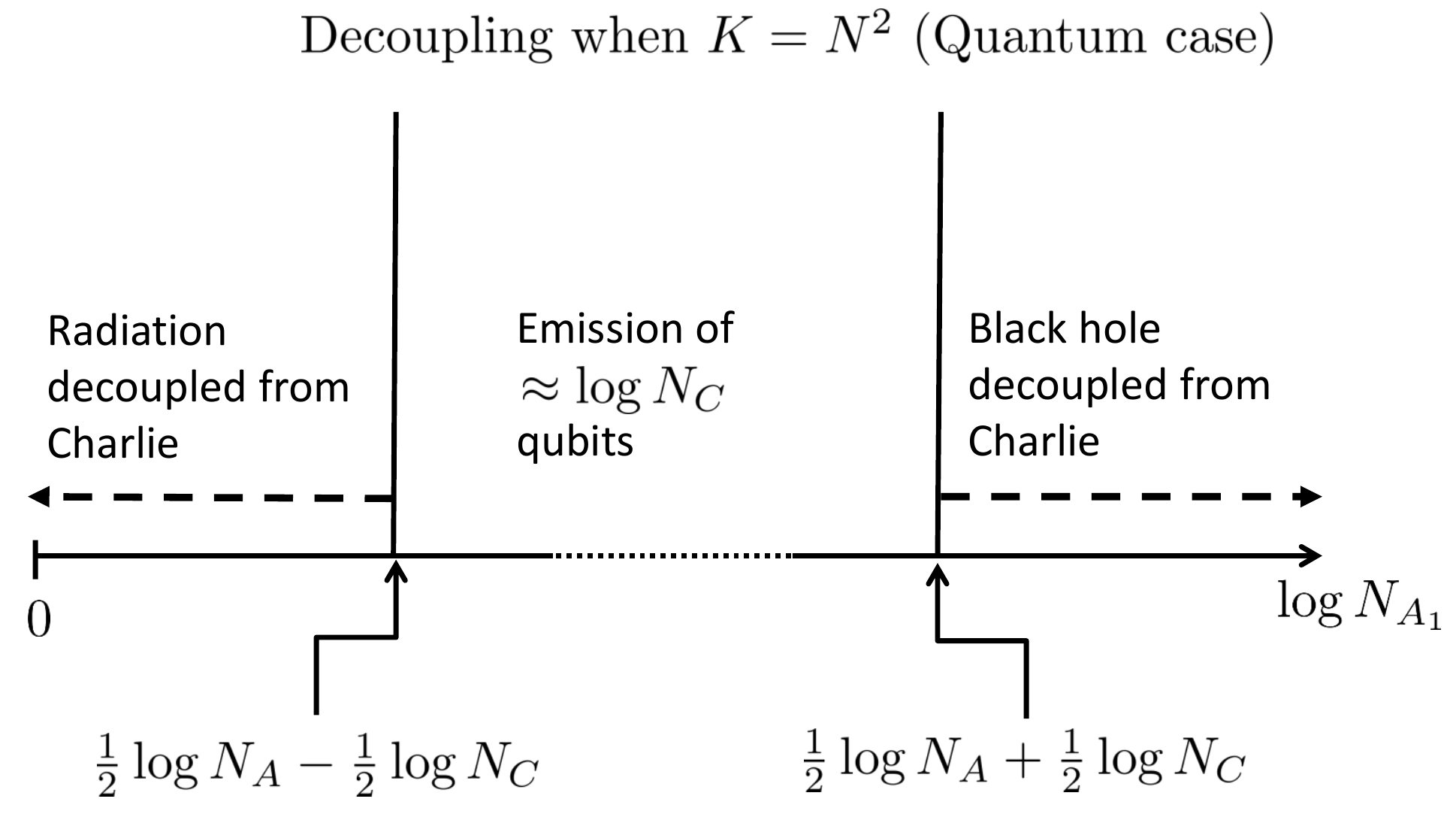}
\caption{In quantum theory, the radiation is initially decoupled from Charlie (information is not in the radiation).  Around the point where half of the black hole's qubits have been radiated away,
this interval ends, and emission of a bit more than $\log N_C$ qubits leads to decoupling of the black hole and Charlie (information not in black hole). Note that the horizontal axis is the number of emitted photons rather than the time coordinate.}
 \label{fig_simp_qt}
 \end{center}
 \end{figure}
In the case of quantum theory (i.e.\ $r=2$), we get the behaviour depicted in Figure~\ref{fig_simp_qt}, which is somehow what we expect.
Shown is the number of generalized bits (in quantum theory, qubits) $\log N_{A_1}$ that have been radiated away since Alice threw her system
into the black hole. 

\begin{figure}[!hbt]
 \begin{center}
\includegraphics[angle=0, width=11cm]{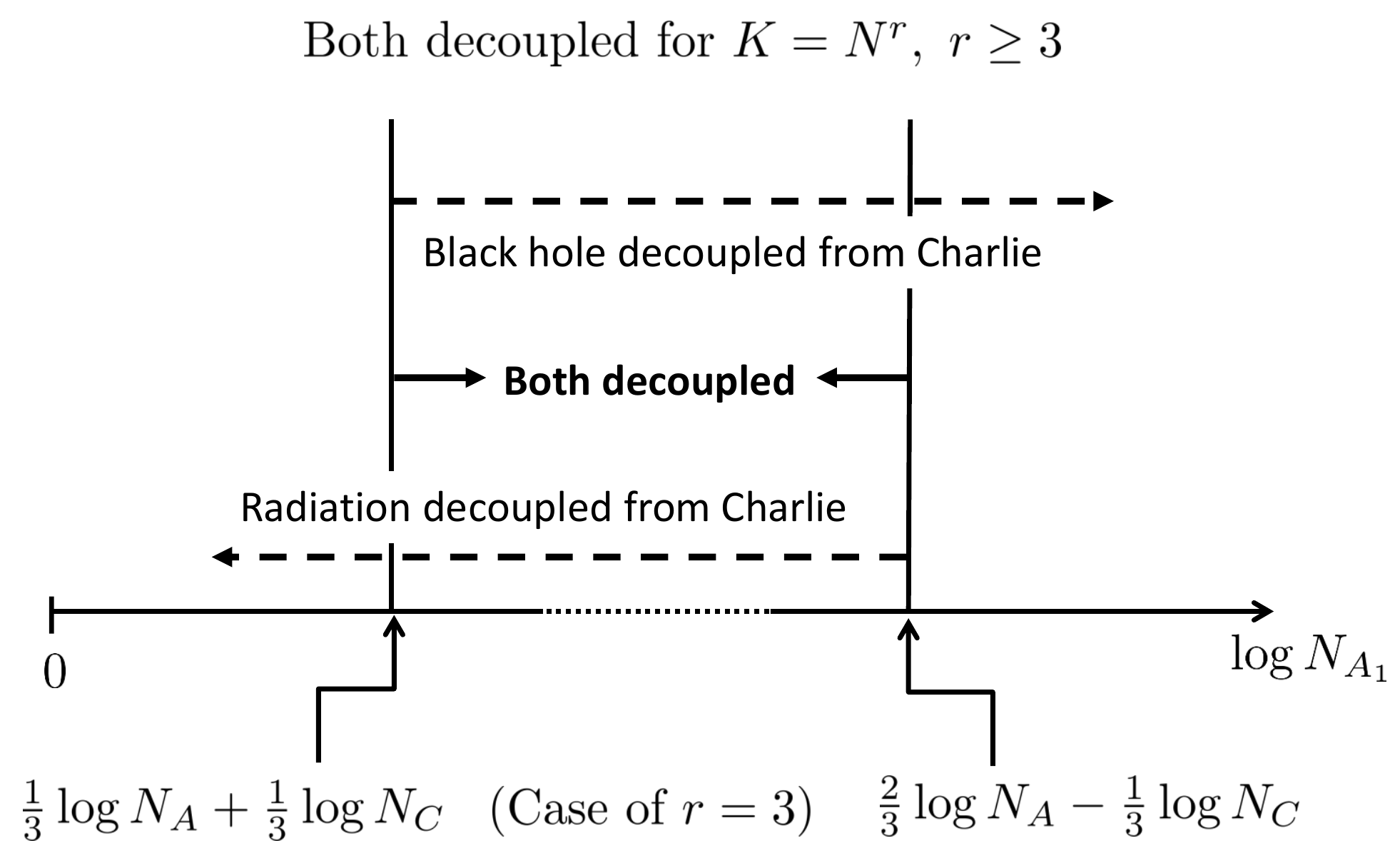}
\caption{Beyond quantum theory, there is a time interval when both the black hole and the radiation are decoupled from Charlie (information is neither in the black hole nor in the radiation).
This shows that there is no analog of quantum theory's ``no-hiding theorem''.}
 \label{fig_simp_post}
 \end{center}
 \end{figure}

In the post-quantum case of $r\geq 3$, the behaviour becomes surprising: we get a time interval in which \emph{both} the black hole and
the radiation are decoupled from Charlie (cf.\ Fig.~\ref{fig_simp_post}). Thus, there cannot be an analog of quantum theory's no-hiding theorem:
decoupling of one system does not guarantee that the information can be extracted on the remaining system. 

\section{Implications for Page's scenario}
We will now consider a further simplification of our scenario, and \textbf{assume that the system $C$ does not exist}, i.e.\ $N_C=K_C=1$.
That is, the black hole is formed from a pure state $\psi^A=\psi^{MB}$ at time $t_0$, and then gradually radiates away. This will allow
us to analyse Page's~\cite{Page} scenario in the context of GPTs. Assuming that the black hole implements a random reversible transformation
(as we do), Page computed the expected entropy of Hawking radiation, i.e.\ the entanglement entropy of the bipartite state, in terms of the
number of emitted quanta.

While there is no unique generalization of Shannon or von Neumann entropy to GPTs~\cite{hein1979entropy, Barnum09, ShortW09}, we can employ our results by
instead considering the \emph{R\'enyi entropy of order $2$}. For a quantum state $\rho$, this is defined as $H_2(\rho):=-\log \tr(\rho^2)$, which
is zero for pure states and has the maximal value $\log n$ for the maximally mixed state on $\C^n$. The expression $\tr(\rho^2)$ equals the purity
$\p(\rho)$ up to some offset and factor. Taking these into account motivates the definition
\[
   H_2(\omega^A):=-\log \left( \frac 1 {N_A} + \frac{N_A-1}{N_A} \p(\omega^A)\right)
\]
for arbitrary state spaces $A$ that satisfy our standard assumptions. This agrees with R\'enyi $2$-entropy in quantum theory by construction;
moreover, it turns out to agree with the classical version of this quantity in classical probability theory. We have $0\leq H_2(\omega^A)\leq \log N_A$
for all state spaces; the minimum is attained iff $\omega^A$ is pure, and the maximum is attained iff $\omega^A=\mu^A$.

\begin{figure}[!hbt]
 \begin{center}
\includegraphics[angle=0, width=8cm]{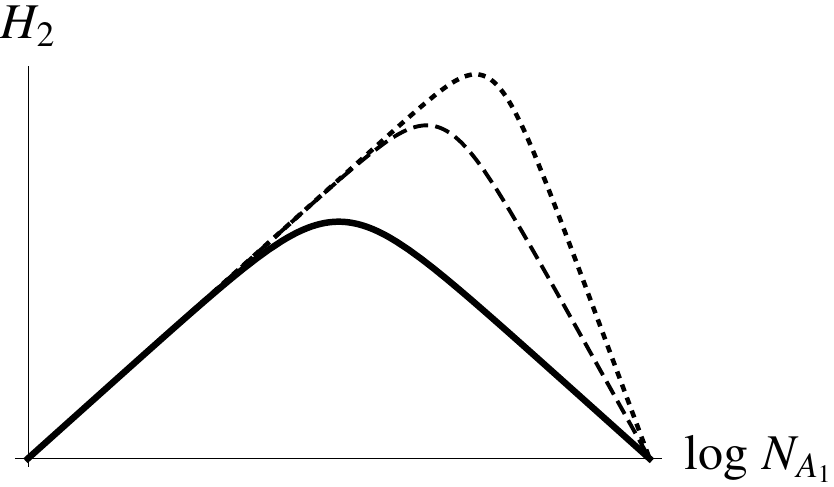} 
\caption{Page's scenario generalised to GPTs. Close to the end of the black hole's lifetime, the Wootters-Hardy parameter $r$ associated with a theory has significant
impact on the rate at which information leaves the hole. The full line is the standard quantum case with $r=2$, reproducing Page's curve~\cite{page-unitary-evap};
the dashed line corresponds to $r=3$, and the dotted line to $r=4$.}
 \label{fig_graph}
 \end{center}
 \end{figure}

According to Theorem~\ref{TheDecoup}, we have
\[
   \int_{U\in \G_A} \p\left(\sigma(U)^{A_1}\right)\, dU = \int_{U\in\G_A} \left\| \sigma(U)^{A_1} - \mu^{A_1}\right\|_2^2 \, dU = \frac{K_{A_1}-1}{N_{A_1}-1}
   \cdot \frac{N_A-1}{K_A-1}.
\]
As usually, in high dimensions, we have a concentration of measure effect, such that in fact for ``almost all'' $U\in\G_A$, we expect
\begin{eqnarray*}
   \p\left(\sigma(U)^{A_1}\right) \approx \frac{N_{A_1}^r -1}{N_{A_1}-1} \cdot \frac{N_A -1}{N_A^r-1}
   \quad\Rightarrow\quad
   H_2\left(\sigma(U)^{A_1}\right)\approx \log N_{A_1}-\log\left(
      1+\frac{(N_{A_1}^r-1)(N_A-1)}{N_A^r-1}
   \right).
\end{eqnarray*}
This gives us an estimate of the Hawking radiation's R\'enyi-$2$ entropy $H_2(A_1)$, depending on the number of emitted degrees of freedom $N_{A_1}$,
or emitted generalised bits $\log N_{A_1}$. Figure~\ref{fig_graph} shows plots of $H_2(A_1)$ over the number of emitted bits
$\log N_{A_1}$ for different $r$'s. For the quantum case, i.e.\ $r=2$, we recover Page's result: the entropy grows
until half of the black hole's qubits have been emitted, and then decreases again. In the post-quantum realm, we see that the entropy keeps growing
for a longer period. In other words, a ``smaller'' post-quantum black hole can purify a ``larger'' amount of outgoing radiation. It reveals information
later in its lifetime than in the quantum case.

Interestingly, the different theories all behave very similarly for small times, but differ strongly among each other (and for $r\neq 2$ from
quantum theory) towards the end of the black hole's lifetime, when quantum gravity effects are expected to dominate.
This implies that in theories which are more general than quantum theory, one can respect the semi-classical result of Hawking (that outgoing radiation appears thermal) for 
much longer times.  And for sufficiently large values of $r$, the black hole can behave semi-classically until the Planck scale, when we anyway expect quantum gravitational effects to occur.
This gives some hope that one can resolve the black hole information problem within the context of theories which preserve information, yet still reproduce low energy phenomena such as
Hawking radiation.  Such theories have interesting properties -- the black hole emits thermal radiation through most of its lifetime, until it approaches Planck size.  At this point,
a small object is purifying a very large system (the emitted radiation).

In quantum theory, such a small object would need to have a huge amount of entropy, violating the conjectured
entropy bound of Bekenstein, and suffering from several issues associated with long-lived remnants.  For example, objects with high entropy take a long time to release
their information, meaning that the black hole would take a large amount of time to finish evaporating~\cite{acn87,CarlitzWilley}.  Highly entropic objects are also expected to 
couple very strongly to all other interactions, as they have a phase space factor which is proportional to the number of degrees of freedom $N$ they possess~\cite{tHooft-bhcompl}.

These problems need not plague the final stages of black hole evaporation in these generalised theories.   At the end of its lifetime, the black hole
will have large $K$, so that it is described by a large number of parameters, but almost all of these states are physically indistinguishable (it has a small value of $N$), 
and so its physical entropy is very small~\cite{MO-gpt-entropy}. In this sense, it would not violate entropy bounds, and the final few bits of information could come out at a reasonably fast rate.
It is unclear whether the black-hole decay rate would have a phase space factor proportional to $N$ or to $K$, a question which has not been considered, given that $N$ and $K$ are so closely related
in quantum and classical theory.  If it were proportional to $N$ rather than $K$, it would mean that Planck sized black holes which purify large systems need not couple strongly to other forms of matter,
evading one of the main objections to such objects.

\section{Concluding remarks}
Given the difficulty encountered when trying to apply standard quantum theory to gravity, it is reasonable to assume that a theory which consistently combines quantum theory and gravity will have to modify quantum theory in some way.  Given the central importance of black holes in guiding our research in finding such a theory, it
is natural to explore how our understanding of black holes changes if we modify quantum theory. The goal of this work was to obtain a glimpse on possible new effects and modified information-theoretic behaviour that might appear in the more general situation beyond quantum theory.

Here, we have seen that several
important aspect of the black information problem are modified in the more general setting.  In particular, if the fundamental theory of nature preserves information, 
than the point at which information is likely to escape a black hole is theory dependent.  For a broad and general class of generalisations of 
quantum theory, we have seen that the point at which information escapes a black hole is given by the Wootters-Hardy parameter $r$, a parameter which relates
the degrees of freedom needed to describe a state with the degrees of freedom which can be distinguished in a measurement.  Quantum theory
corresponds to $r=2$, and the case where information must begin to escape the black hole when half the Hawking photons have escaped.  This makes
it difficult to reconcile unitarity with the apparently semi-classical nature of black hole radiation.
However, by increasing the parameter, one can delay the point at which information generically escapes the black hole, even to the point where it only
escapes in the final burst of radiation, when the black hole is no longer semi-classical, and we expect quantum gravitational effects to come into play.

We have also seen that contrary to the quantum case, information can escape the black hole, but not appear outside of it.  There can be a period of time
when the information is delocalised, providing an alternative to black hole complementarity.   This is particularly relevant when the black hole is initially entangled with the outside world, and information thrown in at this point would generically escape very quickly. 

Our result however comes with important caveats.  Here, we have only considered modifying the theory which governs matter, and have not considered changes to the space-time structure.  A full understanding
of the black hole information problem will presumably require a better understanding of what the quantum theory of gravity will look like.  We also do not know how viable the various generalisations of quantum theory
are; explicitly constructing and classifying these theories is subject of current research.
So, although we may consider theories with larger values of $r$, it may be that these are ruled out by other considerations.  Constructing such theories, and understanding them better, remains an important task.

Finally, it is worth pointing out that our generalised decoupling result gives an amusing insight into quantum information theory.  It has long been a source of discussion as to why
quantum theory differs from classical information theory, often by a factor of $1/2$.  For example, super-dense coding~\cite{BennettWiesner} allows us to send twice as much
information as in the classical case.  Or for the decoupling theorem in the quantum case, we have  $\log{d_{A_1}}\gg  \frac 1 2 \cdot n \cdot I(C:A)$.  Here we see that the $1/2$ in this expression comes from the Wotters-Hardy parameter $r$.
It is because in quantum theory, $\log N$ bits are carried by systems with $2\log K$ parameters and in more general theories, $r \log K$.\\

{\bf Acknowledgments:} We gratefully acknowledge feedback on a draft from Samuel Braunstein, David Jennings, Don Page and Arun Pati.
J.O.\ and M.M.\ are grateful to the Aspen Center for Physics and NSF grant 1066293, where some of this research was completed.
J.O.\ would like to thank the Royal Society for their support. O.D.\ acknowledges support from the 
National Research Foundation (Singapore). O.D.\ was frequently visiting Imperial College whilst undertaking this research.
Research at Perimeter Institute for Theoretical Physics is supported in part by the Government of Canada through NSERC and by the Province of Ontario through MRI.

\bibliography{../refgrav2,../refjono,../refjono2,../refmich2}

\begin{thebibliography}{10}
\bibitem{MO-gpt-entropy}
Modulo certain assumptions, the thermodynamical entropy of any theory is given
  in terms of the states $N$ which can be distinguished by a physical device
  (based on an adaptation of the argument of von Neumann, used to justify the
  fact that the von Neumann entropy has a thermodynamical
  interpretation~\cite{VonNeumann55}). Since the Bekenstein bound is derived
  from thermodynamical considerations, $\log N$ is the entropy which should
  appear in the entropy bound for potential theories of quantum gravity which
  go beyond ordinary quantum theory (M.\ P.\ M\"uller, J.\ Oppenheim, in
  preparation).
\end{thebibliography}
\bibliographystyle{JHEP}

\appendix
\newpage

\section{Technical Supplement: Decoupling Theorem for probabilistic theories}
\subsection{Setup and notation}
\label{SecSetup}

For the main definitions of the GPT framework (together with their physical interpretation), we refer the reader to~\cite{TypEnt}.
We recommend~\cite{Hardy, Barrett07,Mielnik74} as background for readers unfamiliar with GPTs.
We use the definitions and notation as they are introduced in~\cite{TypEnt}, and also some of the results presented there. The GPT
framework contains quantum theory (QT) and classical probability theory (CPT) as  special cases. Table~\ref{TableComparison} gives
an (incomplete) overview.\\

\begin{table}
{\small
\begin{tabular}{|c||c|c|}
\hline
\textbf{General state space $A$} & \textbf{$N$-level QT} & \textbf{$N$-level CPT} \\
\hline\hline
real vector space $A$ & space of $N\times N$ Hermitian matrices & $\R^N$ \\
\hline
$N_A=$max.\ $\#$ of perfectly dist.\ states & Hilbert space dim.\ $N_A=N=\dim\mathcal{H}$ & $N_A=N=\#$ of levels \\
\hline
$K_A=\dim A$, number of parameters & $K_A=N_A^2$, number of real parameters & $K_A=N_A$, number of probabilities \\
to specify unnormalized state & in an unnormalized density matrix & to specify the distribution \\
\hline
$A_+$ set of unnormalized states & positive semidefinite matrices & vectors with non-negative entries \\
\hline
$\Omega_A$ set of normalized states $\omega$ & set of density matrices $\rho$ & set of probability vectors $p$ \\
\hline
$u^A$ unit effect & $u^A(\rho)=\tr(\rho)$ & $u^A(p)=p_1+p_2+\ldots+p_N$ \\
\hline
$\mu^A$ maximally mixed state & $\mu^A=\mathbf{1}/N_A$ & $\mu^A=(1/N_A,\ldots,1/N_A)$ \\
\hline
Bloch vector $\hat\omega\in\hat A$ & $\hat\rho=\rho-\mathbf{1}/N_A$ & $\hat p = (p_1-1/N_A,\ldots, p_{N_A}-1/N_A)$ \\
\hline
invariant inner product $\langle \hat\omega,\hat\varphi\rangle$ & $\langle \hat\omega,\hat\varphi\rangle = \frac {N_A}{N_A-1} \tr(\hat\omega\hat\varphi)$ &
$\langle\hat p,\hat q\rangle=\frac{N_A}{N_A-1}\sum_i \hat p_i \hat q_i$ \\
\hline
purity $\p(\omega)$ & $\p(\rho)=\frac{N_A}{N_A-1}\tr(\rho^2)-\frac 1 {N_A-1}$ & $\p(p)=\frac{N_A}{N_A-1} \sum_i p_i^2 -\frac 1 {N_A-1}$ \\
\hline
group of reversible transf.\ $\G_A$ & projective unitary group, $\rho\mapsto U \rho U^\dagger$ & group of permutations $S_N$ \\
\hline
centered class.\ subsystem $\omega_1,\ldots,\omega_{N_A}$ & ONB $|\psi_1\rangle\langle\psi_1|,\ldots,|\psi_{N_A}\rangle\langle\psi_{N_A}|$ &
$\omega_1=(1,0,\ldots),\ldots,\omega_{N_A}=(\ldots,0,1)$ \\
\hline
$\|\omega-\varphi\|_2$ for normalized states $\omega,\varphi$ & $\|\omega-\varphi\|_2=\sqrt{\frac{N_A}{N_A-1}} \cdot\sqrt{\tr(\omega-\varphi)^2}$ &
$\|\omega-\varphi\|_2=\sqrt{\frac{N_A}{N_A-1}} \cdot\sqrt{\sum_i (\omega_i-\varphi_i)^2}$ \\
\hline
$\|\omega-\varphi\|_1$ for normalized states $\omega,\varphi$ & matrix $1$-norm ($2\times$ trace distance) &
variational distance $\sum_i |\omega_i-\varphi_i|$ \\
\hline
\end{tabular}}
\caption{Our various GPT notions in the special cases of quantum theory (QT) and classical probability theory (CPT).}
\label{TableComparison}
\end{table}

\begin{figure}[!hbt]
 \begin{center}
  \includegraphics[angle=0, width=12cm]{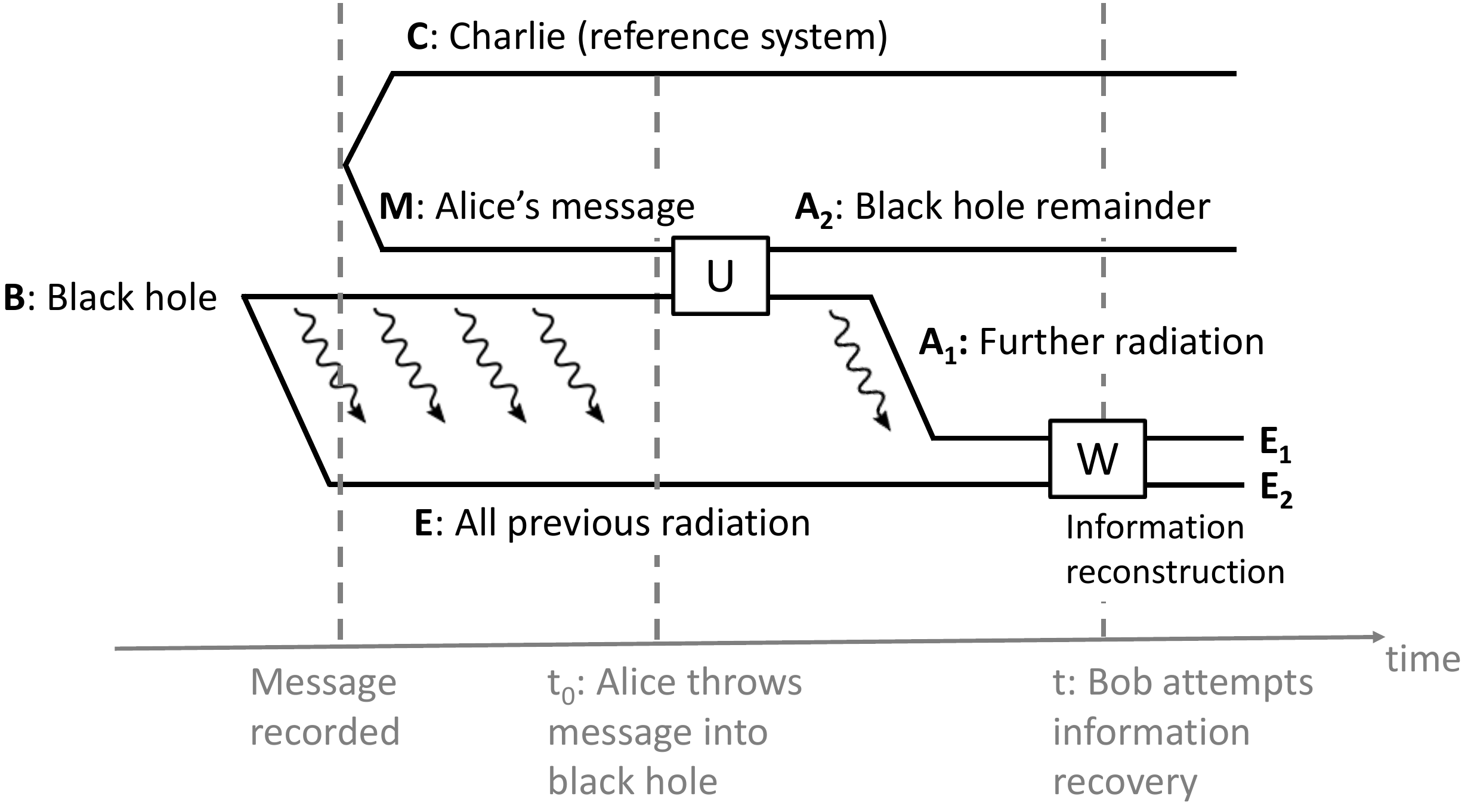}
 \caption{(This figure is also in the main body of the text but is reproduced here in order to make the technical supplement self-contained). Alice throws her message into the black hole. There are three parties: Alice's message (M), the earlier black hole (B), Bob (E) and the reference system (C). (The reference system is particularly natural to include in the case of quantum theory where one may always take C to be a purifying system such that $\psi^{AEC}$ is pure, but we shall not be assuming such purification is always possible.) Then a reversible interaction $U$ is applied to Alice's system, representing the black hole dynamics acting on her diary after she has thrown it in. At some given subsequent time some of the black hole ($A_1$) has leaked out, e.g. via Hawking radiation, and is now in the possession of Bob, and relabeled as $E_2$. Bob also holds any radiation predating Alice's message having entered the black hole ($E$), and he can perform a joint operation $W$ on the system $EA_1=E_1 E_2$. We also have $BM=A_1 A_2 = A$.}   
\label{fig_blackhole_app}
\end{center}
\end{figure}

The quantum decoupling theorem which we shall be generalising is for example described in~\cite{fqsw}. 
The notation is described in Figure~\ref{fig_blackhole_app}.

\subsection{Proof of the GPT decoupling theorem}
When considering more general theories we shall be making certain assumptions which we now state more formally. Most of these
assumptions are necessary if one wants to probe the issue of information escape in black holes, 
and their justification is discussed in the main body of this paper.

\begin{definition}
\label{DefStandardAssumptions}
We say that a collection of state spaces satisfies the \emph{standard assumptions} if the following conditions are all satisfied:
\begin{itemize}
\item \textbf{Transitivity:} For every pair of pure states $\varphi,\omega$ on a state space $A$, there is a reversible
transformation $G\in\G_A$ with $G\varphi=\omega$.
\item \textbf{Local tomography:} If $A$ and $B$ are state spaces, then the joint state space $AB$ has the property
that global states are uniquely determined by the statistics and correlations of local measurement outcomes.
\item \textbf{Centered classical subsystems:} Every state space $A$ contains at least one dynamical centered classical subsystem
$\omega_1,\ldots,\omega_{N_A}$ --- that is, a set of pure and perfectly distinguishable states $\omega_1,\ldots,\omega_{N_A}$ that average to the maximally mixed state,
\[
   \mu^A=\frac 1 {N_A} \sum_{i=1}^{N_A} \omega_i,
\]
such that every permutation on this set of states can be accomplished by some reversible transformation. Moreover,
we assume that composite state spaces $AB$ contain at least one dynamical centered classical subsystem which
is the product of two such classical subsystems on $A$ and $B$.
\item \textbf{Irreducibility:} On every state space $A$ (also the composite ones), the group of reversible transformations 
is an irrep in the usual sense~\cite{Simon}, i.e.\ acts irreducibly on the complexification of $\hat A:=\{x\in A\,\,|\,\, u^A(x)=0\}$.
(Note that this is a bit stronger than the irreducibility in~\cite{TypEnt}), which merely demands that $\G_A$ does not leave any non-trivial subspaces
of the real vector space $\hat A$ invariant).
\end{itemize}
\end{definition}

We shall use certain definitions from~\cite{TypEnt} which generalise key quantum quantities such as purity. We include them here for completeness--for more details such as the operational interpretation of purity see~\cite{TypEnt}. 

\begin{definition}[Maximally mixed state]
\label{DefMaxMix}
If $A$ is a transitive dynamical state space, let $\omega\in\Omega_A$ be an arbitrary pure state, and define
the \emph{maximally mixed state} $\mu^A$ on $A$ by
\[
   \mu^A:=\int_{G\in\G_A} G(\omega)\, dG.
\]
\end{definition}
The Haar measure on the evolution group $G$ exists because G is compact~\cite{Simon}.
Note that it follows from transitivity that $\mu^A$ does not depend on the choice of $\omega$.
\begin{definition} [Bloch vector]
Given any state $\omega\in \Omega_A$
(or, more generally, any point $\omega\in A$ with $u^A(\omega)=1$), we define its corresponding \emph{Bloch vector} $\hat\omega$ as

\[
   \hat\omega:=\omega-\mu^A.
\]
\end{definition}

\begin{definition}[Purity]
\label{DefPurity}
Let $A$ be a transitive and irreducible state space, and let $\langle\cdot,\cdot\rangle$ be the unique
inner product on $\hat A$ such that all transformations are orthogonal and $\langle\hat\alpha,\hat\alpha\rangle=1$
for pure states $\alpha$. Then, the \emph{purity} $\p(\omega)$ of any state $\omega\in\Omega_A$ is defined as
the squared length of the corresponding Bloch vector, i.e.
\[
   \p(\omega):=\|\hat\omega\|^2 \equiv\langle\hat\omega,\hat\omega\rangle.
\]
\end{definition}
 
We have $\p=1$ for pure states and $\p=0$ for the maximally mixed state, cf.~\cite{TypEnt}. Now we prove a claim made in the main text.
\begin{lemma}
\label{LemKNr}
Suppose that we have a family $\mathcal{F}$ of state spaces, such that for every $n\in\mathbb{N}$, there is exactly one state space $A(n)\in\mathcal{F}$ such that
the maximal number of perfectly distinguishable states is $N_{A(n)}=n$. Moreover, suppose that every pair of state spaces $A,B\in\mathcal{F}$ has a composite
$AB\in\mathcal{F}$, and that $m<n$ implies for the state space dimensions $K_{A(m)}<K_{A(n)}$.
If all state spaces in $\mathcal{F}$ and their compositions satisfy our standard assumptions,
then there is an integer $r\in\mathbb{N}$ such that for every state space $A\in\mathcal{F}$,
\[
    K_A=N_A^r.
\]
\end{lemma}
\proof
As we have shown, our standard assumptions imply that $K_{AB}=K_A K_B$ and $N_{AB}=N_A N_B$. Furthermore, the assumptions of the lemma
imply that $K$ is a strictly increasing function of $N$. Thus, it follows from~\cite[Appendix 2]{Hardy} that there is some integer $r\in \mathbb{N}$ such
that $K=N^r$.
\qed

Any state space $A$ can be decomposed into a direct sum of a $(K_A-1)$-dimensional subspace $\hat A$, 
containing all vectors $x$ with $u^A(x)=0$, 
and the one-dimensional subspace $\R\cdot\mu^A$, the span of the maximally mixed state $\mu^A$. So far,
we have introduced an invariant inner product on $\hat A$. It will be convenient for the following calculation
to extend this inner product to all of $A$ in a particular way:
\begin{definition}
If an irreducible state space $A$ carries a centered classical subsystem, then we define an inner product on
$A$ by 
\[
   \langle x,y\rangle:=\langle \hat x,\hat y\rangle+\frac{x_0 y_0}{N_A-1},
\]
where $x$ and $y$ are decomposed as $x=\hat x + x_0 \mu^A$, $y=\hat y+y_0\mu^A$, such that $\hat x,\hat y\in\hat A$.
\end{definition}

The choice of the denominator $(N_A-1)$ seems arbitrary at first sight, but it turns out to simplify calculations a lot.
This is due to the following lemma:
\begin{lemma}
\label{LemHereConditions}
Suppose that $\{A,B,AB\}$ satisfy the standard assumptions.
Then, for arbitrary vectors $\alpha^A,\gamma^A\in A$ and $\beta^B,\delta^B\in B$, we have
\begin{equation}
   \langle\alpha^A\otimes\beta^B|\gamma^A\otimes\delta^B\rangle=\frac{(N_A-1)(N_B-1)}{N_A N_B-1}
   \langle\alpha^A|\gamma^A\rangle\langle\beta^B|\delta^B\rangle.
   \label{eqProdInnerProd}
\end{equation}
In other words,
\[
   \langle\alpha^A\otimes\beta^B|=\frac{(N_A-1)(N_B-1)}{N_A N_B-1}\langle\alpha^A|\otimes\langle\beta^B|.
\]
Moreover, the unit effect $u^B$ on $B$ is $\langle u^B|=(N_B-1)\langle \mu^B|$, that is,
\[
   u^B(x)=(N_B-1)\langle \mu^B|x\rangle\mbox{ for all }x\in B.
\]
\end{lemma}
\proof Recall the decomposition of $AB\equiv A\otimes B$ into the four subspaces
\[
   AB=(\hat A\otimes\hat B)\oplus(\hat A\otimes\mu^B)\oplus(\mu^A\otimes\hat B)\oplus \R\mu^{AB}.
\]
We know that these four subspaces are mutually orthogonal. Therefore, writing any vector $\alpha^A\in A$ as
$\alpha^A=\hat\alpha^A+\alpha_0\mu^A$, where $\hat\alpha^A\in\hat A$ (and similarly for the other vectors), we get
\begin{eqnarray*}
   \langle \alpha^A\otimes\beta^B|\gamma^A\otimes\delta^B\rangle&=&
   \langle \hat\alpha^A\otimes\hat\beta^B+\beta_0 \hat\alpha^A\otimes\mu^B+\alpha_0 \mu^A\otimes \hat\beta^B+\alpha_0\beta_0\mu^{AB}|
   \hat\gamma^A\otimes\hat\delta^B+\delta_0 \hat\gamma^A\otimes \mu^B+\gamma_0 \mu^A\otimes\hat\delta^B+\gamma_0\delta_0\mu^{AB}\rangle\\
   &=&\langle\hat \alpha^A\otimes\hat\beta^B|\hat\gamma^A\otimes\hat\delta^B\rangle + \beta_0\delta_0 \langle\hat\alpha^A\otimes\mu^B|
   \hat\gamma^A\otimes\mu^B\rangle
   +\alpha_0\gamma_0 \langle\mu^A\otimes\hat\beta^B|\mu^A\otimes\hat\delta^B\rangle
   +\alpha_0 \beta_0 \gamma_0 \delta_0 \langle\mu^{AB}|\mu^{AB}\rangle.
\end{eqnarray*}
According to~\cite{TypEnt}, we know the values of all inner products except for the first one: we have $\displaystyle
\langle\hat\alpha^A\otimes\mu^B|\hat\gamma^A\otimes\mu^B\rangle=\frac{N_A-1}{N_A N_B-1} \langle\hat\alpha^A|\hat\gamma^A\rangle$
and $\displaystyle \langle\mu^A\otimes\hat\beta^B|\mu^A\otimes\hat\delta^B\rangle=\frac{N_B-1}{N_A N_B-1} \langle\hat\beta^B|\hat\delta^B\rangle$.
By definition, $\displaystyle\langle\mu^{AB}|\mu^{AB}\rangle=\frac 1 {N_A N_B-1}$. The only unknown term is the first one. Now we argue
by group theory: if $\G_A$ and $\G_B$ act complex-irreducibly on $\hat A$ and $\hat B$, respectively, then $\G_A\otimes\G_B$ acts
complex-irreducibly on $\hat A\otimes \hat B$. Hence the invariant inner product is of the form
$\langle \hat\alpha^A\otimes\hat\beta^B|\hat\gamma^A\otimes\hat\delta^B\rangle=\xi \langle\hat\alpha^A|\hat\gamma^A\rangle
\langle\hat\beta^B|\hat\delta^B\rangle$, with some constant $\xi>0$. We can determine this constant by considering the special case
where $\alpha^A$ and $\beta^B$ are pure states, and $\gamma^A=\alpha^A$ as well as $\delta^B=\beta^B$, then
\begin{eqnarray*}
   \langle\alpha^A\otimes\beta^B|\gamma^A\otimes\delta^B\rangle&=&\langle (\alpha^A\otimes\beta^B)^\wedge+\mu^{AB}|
   (\alpha^A \otimes \beta^B)^\wedge+\mu^{AB}\rangle\\ &=& \p(\alpha^A\otimes\beta^B)+\frac 1 {N_A N_B-1}=1+\frac 1 {N_A N_B-1}.
\end{eqnarray*}
Comparing this with the formula above gives $\displaystyle \xi=\frac{(N_A-1)(N_B-1)}{N_A N_B-1}$. Substituting $\xi$,
we may compare the formula above with the right-hand side of~(\ref{eqProdInnerProd}), and see that they agree.
The second part of the lemma is proven by the calculation
\[
   u^B(x)=u^B(\hat x)+x_0 u^B(\mu^B)=x_0=(N_B-1)\left(\strut \langle\mu^B|\hat x\rangle+x_0\langle\mu^B|\mu^B\rangle\right)
   =(N_B-1)\langle\mu^B|x\rangle,
\]
where have again used the decomposition $x=\hat x+x_0 \mu^B$ with $\hat x\in \hat B$.
\qed

Similar reasoning proves the following lemma:
\begin{lemma}
\label{LemInnerPauli}
If $X^A:A\to\R$ is a linear functional with $X^A(\mu^A)=0$, then there is a vector $\hat X^A\in \hat A$
such that $\langle\hat X^A,v\rangle=X^A(v)$ for all $v\in A$. This vector satisfies
\begin{eqnarray*}
   \langle X^A\otimes u^B|= (N_B-1)\langle\hat X^A|\otimes\langle\mu^B| \qquad\mbox{and}\qquad
   |X^A\otimes u^B\rangle=\frac{N_A N_B-1}{N_A-1}|\hat X^A\rangle\otimes |\mu^B\rangle.
\end{eqnarray*}
\end{lemma}

Twirling over one subsystem produces the maximal mixture on that subsystem. In more detail, we have the following result:
\begin{lemma}
\label{LemTwirling1}
Suppose that $B$ is a transitive state space. Then, for all bipartite states $\omega^{AB}$,
\[
   \int_{T\in\G_B} (\Id\otimes T)(\omega^{AB})dT=\omega^A\otimes\mu^B,
\]
where $\omega^A$ is the local reduced state on $A$.
\end{lemma}
\proof
There are pure states $\varphi_i^A$ spanning $A$, and pure states $\varphi_j^B$ spanning $B$. Hence
their products $\varphi_i^A\otimes\varphi_j^B$ span $AB$, and $\omega^{AB}$ can be written
$\displaystyle\omega^{AB}=\sum_{ij}\alpha_{ij}\varphi_i^A\otimes\varphi_j^B$
with some real and not necessarily non-negative coefficients $\alpha_{ij}\in\R$ and $\sum_{ij}\alpha_{ij}=1$. Hence
\begin{eqnarray*}
\int_{T\in\G_B}(\Id\otimes T)(\omega^{AB})dT&=&\sum_{ij}\alpha_{ij}\int_{T}(\Id\otimes T)(\varphi_i^A\otimes\varphi_j^B)dT
= \sum_{ij} \alpha_{ij} \varphi_i^A\otimes \int_T T\varphi_j^B dT\\
&=& \sum_{ij} \alpha_{ij}\varphi_i^A \otimes \mu^B.
\end{eqnarray*}
But $\sum_{ij} \alpha_{ij}\varphi_i^A$ is the local reduced state $\omega^A$.
\qed

In the following, we need a formula giving the inner product between an arbitrary bipartite state, and a product state
which is maximally mixed on $B$:
\begin{lemma}
Suppose that $\{A,B,AB\}$ satisfy the standard assumptions. Then, for every bipartite state $\alpha^{AB}$,
\[
   \langle \hat\alpha^{AB},(\omega^A\otimes \mu^B)^\wedge\rangle=
   \langle \hat\alpha^A,\hat\omega^A\rangle\cdot
   \frac{N_A-1}{N_A N_B-1},
\]
where $\alpha^A$ is the local reduced state on $A$.
\end{lemma}
\proof
We use twirling and Lemma~\ref{LemTwirling1}:
\begin{eqnarray*}
\langle \hat \alpha^{AB},(\omega^A\otimes\mu^B)^\wedge\rangle&=& \langle \hat \alpha^{AB},\left(
\int_T (\Id\otimes T)(\omega^A\otimes\mu^B)dT\right)^\wedge\rangle
=\langle \hat\alpha^{AB},\int_T (\Id\otimes T)(\omega^A\otimes\mu^B)^\wedge dT\rangle\\
&=& \int_T \langle \hat \alpha^{AB},(\Id\otimes T) (\hat\omega^A\otimes\mu^B) \rangle dT
= \int_T \langle (\Id\otimes T^{-1}) \hat \alpha^{AB},\hat\omega^A\otimes\mu^B \rangle dT\\
&=& \langle\hat\alpha^A\otimes\mu^B,\hat\omega^A\otimes\mu^B\rangle=\langle(\alpha^A\otimes\mu^B)^\wedge,
(\omega^A\otimes\mu^B)^\wedge\rangle.
\end{eqnarray*}
The claim follows.
\qed

\begin{lemma}
\label{LemConditionsRAB}
Suppose that $CA_1 A_2 E$ and all its subsystems satisfy the standard assumptions. Then
\[
   \|\hat\sigma(T)^{C A_2} -(\psi^C \otimes \mu^{A_2})^\wedge \|_2^2 =
   \p(\sigma(T)^{C A_2})-\frac{N_C -1}{N_C N_{A_2} -1}\p(\omega^C)
\]
for every reversible transformation $T\in\G_A$.
\end{lemma}
\proof
We use all the previous lemmas and start by multiplying out the $2$-norm:
\begin{eqnarray*}
   \|\hat\sigma(T)^{C A_2}- (\psi^C \otimes \mu^{A_2})^\wedge\|_2^2&=&
   = \|\hat\sigma(T)^{C A_2}\|_2^2 - 2 \langle \hat\sigma(T)^{C A_2},(\psi^C\otimes\mu^{A_2})^\wedge\rangle
   +\|(\psi^C\otimes\mu^{A_2})^\wedge\|_2^2\\
   &=& \p(\sigma(T)^{C A_2})-2\langle \hat\sigma(T)^C,\hat\psi^C\rangle\cdot \frac{N_C-1}{N_C N_{A_2}-1}+\p(\psi^C\otimes\mu^{A_2}).
\end{eqnarray*}
Finally, use that $\sigma(T)^C=\psi^C$ as well as Theorem 28 from~\cite{TypEnt}.
\qed

The next lemma is a straightforward application of Schur's Lemma. The simple proof is omitted.
\begin{lemma}\label{LemSchur2}
Suppose that $A$ is transitive and complex-irreducible.
Denote by $\Id_{\hat A}$ the orthogonal projector onto $\hat A$ (which is also the identity map on $\hat A$), and by
$\Id_{\mu^A}=(N_A-1)|\mu^A\rangle\langle\mu^A|$ the orthogonal projector onto the span of $\mu^A$. Then, for any matrix $M$ of compatible dimensions, 
\begin{eqnarray*}
   \int_{G\in\G_A} G M G^{-1}\, dG&=&\frac{\Tr(\Id_{\hat A}M\Id_{\hat A})}{K_A-1}\Id_{\hat A}+
   \Tr(\Id_{\mu^A} M \Id_{\mu^A})\Id_{\mu^A}\\
   &=&\frac{\Tr(\Id_{\hat A}M\Id_{\hat A})}{K_A-1}\Id_{\hat A}+
   (N_A-1)\langle\mu^A|M|\mu^A\rangle\Id_{\mu^A}.
\end{eqnarray*}
\end{lemma}
Note that the condition of complex-irreducibility is important here; in the case of real-irreducibility only, the claim of
the lemma would only necessarily be true if $M\geq 0$. Here is an example for how we will use that lemma in the following:
if $\omega$ and $\varphi$ are states on $A$, then
\[
   \int_{G\in\G_A}G |\omega\rangle\langle\varphi| G^{-1}\, dG =\frac{\langle\hat\varphi|\hat\omega\rangle}{K_A-1}\Id_{\hat A}
   +|\mu^A\rangle\langle\mu^A|.
\]

\begin{lemma}
\label{LemI}
If $CA_1 A_2 E$ and all its subsystems satisfy the standard assumptions, then
\begin{eqnarray*}
   I&:=&\int_{T_A\in\G_A}(\Id_C\otimes T_A)|\psi^{CA}\rangle\langle\psi^{CA}|(\Id_C\otimes T_A^{-1})\, dT_A\\
   &=&\left(\Tr_A |\psi^{CA}\rangle\langle\psi^{CA}|\right)\otimes\frac{\Id_{\hat A}}{K_A-1} 
   + \frac{N_C-1}{N_C N_A-1}|\psi^C\rangle\langle\psi^C|\otimes
   \left(\Id_{\mu^A}-\frac{\Id_{\hat A}}{K_A-1}\right).
\end{eqnarray*}
Note that there does not seem to be a simple relation between $\Tr_A|\psi^{CA}\rangle\langle\psi^{CA}|$ and
$|\psi^C\rangle\langle\psi^C|$ in general (the analog of taking the partial trace in quantum theory would instead
be ${\rm Id}_C\otimes u^A(\psi^{CA})=\psi^C$).
\end{lemma}
\proof
There are real numbers $w_{ij}$, summing to one, and pure states $\phi_i^C\in\Omega_C$ and $\phi_j^A\in\Omega_A$ such that
$\displaystyle\omega=\sum_{i=1}^{K_C}\sum_{j=1}^{K_A} w_{ij}\phi_i^C\otimes\phi_j^A$.
Then, $I=\sum_{ijkl}w_{ij}w_{kl}I_{ijkl}$, with
\begin{eqnarray*}
   I_{ijkl}&=&\int_{T_A\in\G_A}(\Id_C\otimes T_A)|\phi_i^C\otimes\phi_j^A\rangle\langle\phi_k^C\otimes\phi_l^A|(\Id_C\otimes T_A^{-1})\,dT_A\\
   &=&\int_{T_A\in\G_A}(\Id_C\otimes T_A)|\phi_i^C\rangle\otimes|\phi_j^A\rangle\langle\phi_k^C|\otimes\langle\phi_l^A|(\Id_C\otimes T_A^{-1})\,dT_A
    \cdot \frac{(N_C-1)(N_A-1)}{N_C N_A-1}\\
   &=&\frac{(N_C-1)(N_A-1)}{N_C N_A-1} \int_{T_A\in\G_A}|\phi_i^C\rangle\langle\phi_k^C|\otimes T_A|\phi_j^A\rangle\langle\phi_l^A|T_A^{-1}\,dT_A\\
   &=&\frac{(N_C-1)(N_A-1)}{N_C N_A-1} |\phi_i^C\rangle\langle\phi_k^C|\otimes\int_{T_A\in\G_A} T_A|\phi_j^A\rangle\langle\phi_l^A|T_A^{-1}\,dT_A\\
   &=&\frac{(N_C-1)(N_A-1)}{N_C N_A-1} |\phi_i^C\rangle\langle\phi_k^C|\otimes
   \left(
      \frac{\langle\hat\phi_l^A|\hat\phi_j^A\rangle}{K_A-1}\Id_{\hat A}+|\mu^A\rangle\langle\mu^A|
   \right).
\end{eqnarray*}
Then, substitute the identity $\displaystyle\langle\hat\phi_l^A|\hat\phi_j^A\rangle=\langle\phi_l^A|\phi_j^A\rangle-\frac 1 {N_A-1}$
and multiply out. Put in the sum over $ijkl$ again. The resulting terms can be identified as follows: From
\[
   |\psi^{CA}\rangle\langle\psi^{CA}|=\frac{(N_C-1)(N_A-1)}{N_C N_A-1}\sum_{ijkl} w_{ij} w_{kl} |\phi_i^C\rangle\langle\phi_k^C|
   \otimes |\phi_j^A\rangle\langle\phi_l^A|
\]
it follows that
\begin{eqnarray}
   \Tr_A|\psi^{CA}\rangle\langle\psi^{CA}|= \frac{(N_C-1)(N_A-1)}{N_C N_A-1} 
   \sum_{ijkl} w_{ij} w_{kl} \langle\phi_l^A|\phi_j^A\rangle |\phi_i^C\rangle\langle\phi_k^C|,\label{EqPartialTrace}
\end{eqnarray}
and
\begin{eqnarray*}
   |\psi^C\rangle\langle\psi^C|&=& \left| \sum_{ij} w_{ij} \phi_i^C\right\rangle\left\langle \sum_{kl} w_{kl}\phi_k^C\right|
   = \sum_{ijkl} w_{ij} w_{kl} |\phi_i^C\rangle\langle\phi_k^C|.
\end{eqnarray*}
This proves the claim.
\qed

Using all the previous lemmas, we can express the desired integral over the purity in the following way:
\begin{lemma}
\label{LemJ}
Suppose that $CA_1 A_2 E$ and all its subsystems satisfy the standard assumptions.
If $X^{CA_2}$ is any Pauli map on $CA_2$, then
\begin{equation}
   \int_{T\in\G_A} \p(\sigma(T)^{C A_2})\,dT=(K_C K_{A_2}-1)\langle X^{CA_2}\otimes u^{A_1}|J|X^{CA_2}\otimes u^{A_1}\rangle,
   \label{eqRandomPurity}
\end{equation}
where
\begin{equation}
    J:=\int_{G\in\G_{CA_2}} (G\otimes\Id_{A_1})I(G^{-1}\otimes\Id_{A_1})\, dG,
    \label{EqJ}
\end{equation}
and $I$ is the expression from Lemma~\ref{LemI}.
\end{lemma}
\proof
As shown in~\cite{TypEnt}, purity can be expressed as a twirling integral:
\[
   \p(\sigma(T)^{C A_2})=(K_C K_{A_2}-1)\int_{G\in\G_{CA_2}} \left((X^{CA_2}\circ G)(\sigma(T)^{CA_2})\right)^2\, dG,
\]
if $X^{CA_2}$ is any Pauli map on $CA_2$. Substituting
\begin{eqnarray*}
   X^{CA_2}\circ G(\sigma(T)^{CA_2})&=&X^{CA_2}\otimes u^{A_1}\left(G\otimes\Id_{A_1}(\sigma(T)^{CA})\right)
   = \langle X^{CA_2}\otimes u^{A_1}|G\otimes\Id_{A_1}|\sigma(T)^{CA}\rangle
\end{eqnarray*}
and $\sigma(T)^{CA}=\Id_C\otimes T_A \psi^{CA}$
and then reversing the order of integration proves the claim.
\qed

Due to Schur's Lemma, evaluating twirling integrals amounts to projecting onto invariant subspaces. Therefore, we need
the following lemma that states some of the projection identities that will be used in this procedure.
\begin{lemma}
\label{LemProjections}
Under the standard assumptions, we have the following projection identities:
\begin{eqnarray*}
   \Id_{\mu^{CA_2}} \left(\strut |\psi^C\rangle\langle\psi^C|\otimes\Id_{A_2}\right)\Id_{\mu^{CA_2}}&=& \frac{\Id_{\mu^{CA_2}}}{N_C-1},\\
   \Id_{\mu^{CA_2}} \left(\strut |\psi^C\rangle\langle\psi^C|\otimes\Id_{\mu^{A_2}}\right)\Id_{\mu^{CA_2}}&=& \frac{\Id_{\mu^{CA_2}}}{N_C-1},\\
   \Id_{(CA_2)^\wedge} \left(\strut |\psi^C\rangle\langle\psi^C|\otimes\Id_{A_2}\right)\Id_{(CA_2)^\wedge}&=&
      |\psi^C\rangle\langle\psi^C|\otimes\Id_{\hat A_2} 
       +  |\hat\psi^C\rangle\langle\hat\psi^C|\otimes\Id_{\mu^{A_2}}\\
    \Id_{(CA_2)^\wedge} \left(\strut |\psi^C\rangle\langle\psi^C|\otimes\Id_{\mu^{A_2}}\right)\Id_{(CA_2)^\wedge}&=&
       |\hat\psi^C\rangle\langle\hat\psi^C|\otimes\Id_{\mu^{A_2}},\\
    \Id_{\mu^{CA_2}} \left(\strut \Tr_A|\psi^{CA}\rangle\langle\psi^{CA}|\otimes\Id_{A_2}\right)\Id_{\mu^{CA_2}}&=&
       \frac{N_A-1}{N_C N_A-1} \langle\psi^A|\psi^A\rangle \Id_{\mu^{CA_2}},\\
    \Id_{\mu^{CA_2}} \left(\strut \Tr_A|\psi^{CA}\rangle\langle\psi^{CA}|\otimes\Id_{\mu^{A_2}}\right)\Id_{\mu^{CA_2}}&=&
       \frac{N_A-1}{N_C N_A-1} \langle\psi^A|\psi^A\rangle \Id_{\mu^{CA_2}},\\
    \Id_{(CA_2)^\wedge}\left(\Tr_A|\psi^{CA}\rangle\langle\psi^{CA}|\otimes \Id_{A_2}\right)\Id_{(CA_2)^\wedge}&=&
       \left(\Tr_A|\psi^{CA}\rangle\langle\psi^{CA}|\right)\otimes\Id_{\hat A_2}\\ &&+
         \left(\Tr_A|\hat \psi^{CA}\rangle\langle\hat\psi^{CA}|\right)\otimes\Id_{\mu^{A_2}},\\
     \Id_{(CA_2)^\wedge}\left(\Tr_A|\psi^{CA}\rangle\langle\psi^{CA}|\otimes \Id_{\mu^{A_2}}\right)\Id_{(CA_2)^\wedge}&=&
       \left(\Tr_A|\hat\psi^{CA}\rangle\langle\hat\psi^{CA}|\right)\otimes\Id_{\mu^{A_2}}.
\end{eqnarray*}
\end{lemma}
\proof
To prove the first four identities,
multiply out the expression in between the two projection operators, using $\psi^C=\hat\psi^C+\mu^C$ and
$\Id_{A_2}=\Id_{\hat A_2}+\Id_{\mu_{A_2}}$, and check for each addend the corresponding action on the
subspace structure.

For the remaining identities, note that due to~(\ref{EqPartialTrace}) we can write
\begin{equation}
   \Tr_A|\psi^{CA}\rangle\langle\psi^{CA}|= \frac{(N_C-1)(N_A-1)}{N_C N_A -1}
   \sum_{ik} \xi_{ik}|\phi_i^C\rangle\langle\phi_k^C|,
   \label{eqPartialTraceDev}
\end{equation}
where $\xi_{ik}=\sum_{jl} w_{ij} w_{kl} \langle \phi_l^A|\phi_j^A\rangle$. Hence
\begin{eqnarray*}
   \sum_{ik}\xi_{ik}&=& \sum_{ijkl} w_{ij} w_{kl} \langle\phi_l^A|\phi_j^A\rangle 
   = \left(\sum_{kl} w_{kl} \langle\phi_l^A|\right)\left(\sum_{ij} w_{ij} |\phi_j^A\rangle\right)
   =\langle\psi^A|\psi^A\rangle 
   = \langle\hat\psi^A|\hat\psi^A\rangle+\langle\mu^A|\mu^A\rangle
   =\p(\psi^A)+\frac 1 {N_A-1}.
\end{eqnarray*}
To compute the remaining four identities, treat the $|\phi_i^C\rangle\langle\phi_k^C|$ appearing in~(\ref{eqPartialTraceDev})
exactly as $|\psi^C\rangle\langle\psi^C|$ in the first four equations.
\qed

Now we are finally ready to prove a first version of the decoupling theorem:
\begin{lemma}
\label{TheDecouplingPurity}
If the standard assumptions are satisfied, then
\begin{eqnarray*}
\int_{T\in\G_A} \p\left(\sigma(T)^{CA_2}\right)dT&=& \p(\psi^{CA})\cdot\frac{(K_{A_2}-1)(N_C N_A-1)}{(N_C N_{A_2}-1)(K_A-1)}
+\p(\psi^C)\cdot\frac{(N_C-1)(K_A-K_{A_2})}{(N_C N_{A_2}-1)(K_A-1)}.
\end{eqnarray*}
\end{lemma}
\proof
We use the identities and notation from Lemma~\ref{LemJ}, and start by computing $J$ using~(\ref{EqJ}). First, we expand $I$
into a sum of tensor product expressions:
\begin{eqnarray*}
   I&=& \left(\Tr_A|\psi^{CA}\rangle\langle\psi^{CA}|\right)\otimes\frac{\Id_{A_1}}{K_A-1}\otimes\Id_{A_2}
   -\left(\Tr_A|\psi^{CA}\rangle\langle\psi^{CA}|\right)\otimes\frac{\Id_{\mu^{A_1}}}{K_A-1}\otimes\Id_{\mu^{A_2}}
   +\frac{N_C-1}{N_C N_A-1}|\psi^C\rangle\langle\psi^C|\otimes\Id_{\mu^{A_1}}\otimes\Id_{\mu^{A_2}} \\
    &&-\frac{N_C-1}{N_C N_A-1}|\psi^C\rangle\langle\psi^C|\otimes\frac{\Id_{A_1}}{K_A-1}\otimes\Id_{A_2}
   +\frac{N_C-1}{N_C N_A-1}|\psi^C\rangle\langle\psi^C|\otimes\frac{\Id_{\mu^{A_1}}}{K_A-1}\otimes\Id_{\mu^{A_2}}.
\end{eqnarray*}
Then, we use Lemma~\ref{LemProjections} and Lemma~\ref{LemSchur2} to twirl over that expression and get $J$.
This involves lengthy, but straightforward algebra und the identity $\Tr\left(\Tr_A|\psi^{CA}\rangle\langle\psi^{CA}|\right)=
\langle\psi^{CA}|\psi^{CA}\rangle$. We first get the following expression for $J$ (some addends are marked for later reference):
\begin{eqnarray*}
   J&=&\frac{\Id_{A_1}}{K_A-1}\otimes \left( \frac{\langle\psi^{CA}|\psi^{CA}\rangle(K_{A_2}-1)+\langle\hat\psi^{CA}|\hat\psi^{CA}\rangle}
   {K_C K_{A_2}-1} \Id_{(CA_2)^\wedge}+\underbrace{\frac{N_A-1}{N_C N_A-1}\langle\psi^A|\psi^A\rangle\Id_{\mu^{CA_2}}}_{(*)}\right)\\
   && - \frac{\Id_{\mu^{A_1}}}{K_A-1}\otimes\left(\frac{\langle\hat\psi^{CA}|\hat\psi^{CA}\rangle}{K_C K_{A_2}-1} \Id_{(CA_2)^\wedge}
   + \underbrace{ \frac{N_A -1}{N_C N_A-1}\langle\psi^A|\psi^A\rangle\Id_{\mu^{CA_2}}}_{(*)}\right)\\
    &&+\frac{N_C-1}{N_C N_A-1} \Id_{\mu^{A_1}}\otimes\left( \frac{\langle\hat\psi^C|\hat\psi^C\rangle}{K_C K_{A_2}-1} \Id_{(CA_2)^\wedge}
   +\underbrace{ \frac{\Id_{\mu^{CA_2}}}{N_C-1}}_{(*)}\right)\\
   && - \frac{N_C-1}{N_C N_A-1}\frac{\Id_{A_1}}{K_A-1} \otimes\left( \frac{\langle\psi^C|\psi^C\rangle(K_{A_2}-1)+
   \langle\hat\psi^C|\hat\psi^C\rangle}{K_C K_{A_2}-1} \Id_{(CA_2)^\wedge}+\underbrace{\frac{\Id_{\mu^{CA_2}}}{N_C-1}}_{(*)}\right)\\
   && + \frac{N_C-1}{N_C N_A-1}\frac{\Id_{\mu^{A_1}}}{K_A-1}\otimes\left( \frac{\langle\hat\psi^C|\psi^C\rangle}{K_C K_{A_2} -1}
   \Id_{(C A_2)^\wedge}+\underbrace{ \frac{\Id_{\mu^{CA_2}}}{N_C-1}}_{(*)}\right).
\end{eqnarray*}
Recall identity~(\ref{eqRandomPurity}). In this identity, we are free to choose an arbitrary Pauli map $X^{CA_2}$ on $CA_2$.
A useful choice is $\displaystyle X^{CA_2}:=\sqrt{\frac{N_C-1}{N_C N_{A_2}-1}} X^C\otimes u^{A_2}$, where $X^C$ is any
Pauli map on $C$. The expression $X^{CA_2}\otimes u^{A_1}$ appearing in~(\ref{eqRandomPurity}) becomes $X^C\otimes u^A$.
In the usual decomposition of $CA$ into four invariant orthogonal subspaces, this functional is zero on all subspaces except
for the subspace $\hat C\otimes \mu^A$. In particular, this functional gives zero if it is evaluated on all addends that are marked
with a $(*)$ above. Therefore, we have $\langle X^C\otimes u^A|J|X^C\otimes u^A\rangle=\langle X^C\otimes u^A|J'|X^C\otimes u^A\rangle$,
where $J'$ is $J$ without the marked addends, that is,
\begin{eqnarray*}
   J'&=& \frac{\langle\psi^{CA}|\psi^{CA}\rangle(K_{A_2}-1)+\langle\hat\psi^{CA}|\hat\psi^{CA}\rangle}{(K_A-1)(K_C K_{A_2}-1)}
   \Id_{A_1}\otimes\Id_{(CA_2)^\wedge}  
   -\frac{\langle\hat\psi^{CA}|\hat\psi^{CA}\rangle}{(K_A-1)(K_C K_{A_2}-1)} \Id_{\mu^{A_1}}\otimes\Id_{(CA_2)^\wedge} \\
    &&+ \frac{(N_C-1)\langle\hat\psi^C|\hat\psi^C\rangle}{(N_C N_A-1)(K_C K_{A_2}-1)} \Id_{\mu^{A_1}}\otimes\Id_{(CA_2)^\wedge}
   +\frac{N_C-1}{(N_C N_A-1)(K_A-1)}\cdot\frac{\langle\hat\psi^C|\hat\psi^C\rangle}{K_C K_{A_2}-1}\Id_{\mu^{A_1}}\otimes\Id_{(CA_2)^\wedge}\\
   &&-\frac{N_C-1}{(N_C N_A-1)(K_A-1)} \cdot\frac{\langle\psi^C|\psi^C\rangle(K_{A_2}-1)+\langle\hat\psi^C|\hat\psi^C\rangle}
   {K_C K_{A_2}-1} \Id_{A_1}\otimes \Id_{(CA_2)^\wedge}.
\end{eqnarray*}
Sandwiching this expression with $X^C\otimes u^A$ removes all other addends that are not supported on $\hat C\otimes\mu^A$. That is,
we can replace once again $J'$ by another expression $J''$ where all those terms are removed. So we have
$\langle X^C\otimes u^A|J|X^C\otimes u^A\rangle=\langle X^C\otimes u^A|J''|X^C\otimes u^A\rangle$, and some simplification yields
\begin{eqnarray*}
   J''&=&\Id_{\hat C\otimes\mu^A} \underbrace{\left(
      \p(\psi^{CA})\frac{K_{A_2}-1}{(K_A-1)(K_C K_{A_2}-1)}+\p(\psi^C) \frac{(N_C-1)(K_A -K_{A_2})}{(N_C N_A-1)(K_C K_{A_2}-1)(K_A-1)}
   \right)}_{=:\xi}.
\end{eqnarray*}
For the final step, we use Lemma~\ref{LemInnerPauli} and calculate
\begin{eqnarray*}
   \int_{T\in\G_A} \p(\sigma(T)^{CA_2})\, dT &=& (K_C K_{A_2}-1)\langle X^{CA_2}\otimes u^{A_1}|J''|X^{CA_2}\otimes u^{A_1}\rangle
   = (K_C K_{A_2}-1) \langle X^{CA_2}\otimes u^{A_1}|X^{CA_2}\otimes u^{A_1}\rangle \xi\\
   &=& (K_C K_{A_2}-1) \xi (N_{A_1}-1)\langle\hat X^{CA_2}|\otimes\langle\mu^{A_1}|\frac{N_{CA_2}N_{A_1}-1}{N_{CA_2}-1}
   |\hat X^{CA_2}\rangle\otimes|\mu^{A_1}\rangle\\
   &=& (K_C K_{A_2}-1)\xi\frac{N_C N_A -1}{N_C N_{A_2}-1}.
\end{eqnarray*}
Resubstituting $\xi$ proves the claim.
\qed

\begin{corollary}[Decoupling, $2$-norm version]
The standard assumptions imply
\begin{eqnarray*}
   \int_{T\in\G_A} \left\| \hat\sigma(T)^{CA_2}-(\psi^C\otimes\mu^{A_2})^\wedge \right\|_2^2\, dT&=&
\p(\psi^{CA})\cdot\frac{(K_{A_2}-1)(N_C N_A-1)}{(N_C N_{A_2}-1)(K_A-1)} 
-\p(\psi^C)\cdot\frac{(N_C-1)(K_{A_2}-1)}{(N_C N_{A_2}-1)(K_A-1)}.
\end{eqnarray*}
\end{corollary}
\proof
Substitute the result of Lemma~\ref{LemConditionsRAB} into Theorem~\ref{TheDecouplingPurity}.
\qed

For technical reasons, we have thus far worked with the $2$-norm distance, $\|\cdot\|_2$. Similarly as in quantum theory, this norm is
easy to work with, because it is invariant with respect to reversible transformations. However -- again, in analogy to quantum theory -- it does not in
itself possess a natural operational interpretation.

It turns out that one additional assumption allows us to relate the $2$-norm to the maximal probability of distinguishing two states, which is the
GPT analog of quantum theory's $1$-norm, or trace distance.

\begin{definition}[$1$-norm]
For any state space $A$, and any vector $x\in A$, define its $1$-norm by
\[
   \|x\|_1:=2 \max_{0\leq e \leq u^A} |e(x)|,
\]
where the maximum is over all effects $e$ with $0\leq e\leq u^A$, i.e.\ $0\leq e(\omega)\leq 1$ for all $\omega\in\Omega_A$.
\end{definition}
Consequently, the $1$-norm of the difference of two states tells us the maximal difference of outcome probabilities of
any possible measurement that can be applied to the states:
\[
   \frac 1 2 \|\varphi-\omega\|_1 = \max_{0\leq e \leq u^A} |e(\varphi)-e(\omega)|\qquad (\varphi,\omega\in\Omega_A).
\]
We would like to relate the $2$-norm and the $1$-norm. The following assumption turns out to be crucial for this.

\begin{definition}
We say that a state space is \emph{bit-symmetric}, if for every two pairs $\varphi,\omega$ and $\varphi',\omega'$
of perfectly distinguishable states, there is a reversible transformation $G$ such that $G\varphi=\varphi'$ and
$G\omega=\omega'$.
\end{definition}

Bit symmetry is a significant assumption but it has a nice physical interpretation, see~\cite{BitSymmetry}.
It allows to bound the $1$-norm in terms of the $2$-norm in a way which is analogous to quantum theory:
\begin{theorem}
\label{TheIneqNorms}
Suppose that $A$ is any bit-symmetric state space which satisfies the standard assumptions. Then
\[
   \|\varphi-\omega\|_1\leq \sqrt{N_A} \|\varphi-\omega\|_2
\]
for all $\varphi,\omega\in\Omega_A$.
\end{theorem}
\proof
In~\cite{BitSymmetry}, it is shown that there is an inner product $[\cdot,\cdot]$ on $A$ with $[\omega,\omega]=1$ for all
pure states $\omega$, and $[\omega,\varphi]=0$ if $\omega$ is pure and $\omega,\varphi$ are perfectly distinguishable.
We give it unusual brackets, because it is in general different to the inner product $\langle\cdot,\cdot\rangle$ that we
have used before in this paper (though closely related). Decomposing vectors $x,y\in A$ as $x=x_0 \mu^A+\hat x$
with $\hat x\in \hat A$ (and similarly for $B$), this inner product satisfies
\[
   [x,y]=\lambda x_0 y_0+(1-\lambda)\langle \hat x,\hat y\rangle
\]
for some $\lambda\in(0,1)$; now, $\langle\cdot,\cdot\rangle$ is our usual inner product on Bloch vectors. Now let
$\omega_1,\ldots,\omega_{N_A}$ be a dynamical centered classical subsystem. The states $\omega_1$ and $\omega_2$
are perfectly distinguishable; from~\cite{TypEnt}, we know that $\langle\hat\omega_1,\hat\omega_2\rangle=-1/(N_A-1)$.
Thus $0=[\omega_1,\omega_2]=\lambda+(1-\lambda)\langle\hat\omega_1,\hat\omega_2\rangle=\lambda-(1-\lambda)/(N_A-1)$,
which determines $\lambda=1/N_A$.

Since the order unit $u^A$ as a vector is invariant with respect to all reversible transformations, it must be a multiple of the
maximally mixed state, $u^A=c\mu^A$ with some $c>0$. For the rest of this proof, let $\|x\|_2:=\sqrt{[x,x]}$ denote the norm
which is derived from this new inner product. Then we have
\[
   \|u^A\|_2^2 = [u^A,u^A]=[u^A,c\mu^A]=c[u^A,\mu^A]=c u^A(\mu^A)=c.
\]
On the other hand,
\[
   \| u^A \|_2^2 = [u^A,u^A] = c^2 [\mu^A,\mu^A] = c^2 \cdot \lambda=c^2/N_A.
\]
This implies that $\|u^A\|_2=\sqrt{N_A}$. Now use the self-duality of $A$, which implies~\cite{Bellissard} that there is a decomposition $\varphi-\omega=R-S$
into effects $R,S\geq 0$ which are orthogonal: $[R,S]=0$. Since $0=u(\varphi-\omega)=u(R-S)$, we have $u(R)=u(S)$. Clearly,
$\omega_R:=R/u(R)$ and $\omega_S:=S/u(S)=S/u(R)$ are normalized states. Thus,
\begin{eqnarray*}
   \frac 1 2 \|\varphi-\omega\|_1&=&\max_{0\leq E \leq u} E(\varphi-\omega)=\max_{0\leq E \leq u} E(R-S)
   =u(R)\max_{0\leq E \leq u} \left(
      E(\omega_R)-E(\omega_S)
   \right)
   \leq u(R) = \frac 1 2 (u(R)+u(S))  \\
   &=& \frac 1 2 u(R+S) \leq \frac 1 2 \|u\|_2 \cdot \|R+S\|_2 = \frac 1 2 \sqrt{N_A} \sqrt{\|R\|_2^2 + \|S\|_2^2} = \frac 1 2 \sqrt{N_A} \|R-S\|_2
   = \frac 1 2 \sqrt{N_A} \|\varphi-\omega\|_2.
\end{eqnarray*}
\qed

It is easy to see that the inequality $\|\varphi-\omega\|_1\leq\sqrt{N_A}\|\varphi-\omega\|_2$ is false in general if the state space is not bit-symmetric
or does not satisfy the standard assumptions -- in particular, the inequality does
not follows from transitivity alone.
As a simple counterexample, imagine a state space which is
a $d$-dimensional cube. It is easy to see that this has $N=2$ distinguishable states (it is a generalized bit); moreover,
it satisfies all standard assumptions (as a stand-alone state space without composites).

Consider two states $\omega, \varphi$ that are adjacent (i.e.\ neighboring) pure states of the $d$-cube. To compute the distance,
we have to imagine that the cube is inscribed into a unit ball. Then the two states have Euclidean distance
$\|\varphi-\omega\|_2=2/\sqrt{d}$, but they are perfectly distinguishable, i.e.\ $\frac 1 2\|\varphi-\omega\|_1=1$.
Thus, in this example, $\|\varphi-\omega\|_1=\sqrt{d}\|\varphi-\omega\|_2$, where $d$ can be arbitrarily large, while $N=2$
is fixed.

Employing the above inequality on the generalised decoupling theorem we then obtain the following decoupling theorem.

\begin{theorem}[Decoupling, $1$-norm version]
\label{TheDec1}
If the subsystem $CA_2$ is additionally bit-symmetric, then 
\begin{eqnarray*}
\int_{U\in\G_A} \left\| \sigma(U)^{CA_2}-\psi^C\otimes\mu^{A_2} \right\|_1^2\, dU&\leq&
\frac{N_C N_{A_2}(K_{A_2}-1)}{(N_C N_{A_2}-1)(K_A-1)} \left[\strut
   \p(\psi^{CA}(N_C N_A-1) -\p(\psi^C)(N_C-1)
\right]\\
&\stackrel < \approx & \p(\psi^{CA})\cdot\frac{N_C N_A}{K_{A_1}} \qquad\mbox{ if all }N,K\mbox{ large.}
\end{eqnarray*}
\end{theorem}

\end{document}